\documentclass[aps,superscriptaddress,twocolumn]{revtex4}
\usepackage{graphicx}
\usepackage{dcolumn}
\usepackage{bm}
\usepackage{diagbox}
\usepackage{multirow}
\usepackage{color}
\usepackage[utf8]{inputenc}
\usepackage[T1]{fontenc}
\usepackage{amsfonts}
\usepackage{amssymb}

\usepackage{lineno,hyperref}    

\newcommand{\MC}[1]{\textcolor{black}{{#1}}}
\newcommand{\CT}[1]{\textcolor{black}{{#1}}}
\newcommand{\Mark}[1]{\textcolor{black}{{#1}}}

\begin{document}

\title{Modeling the High-Pressure Solid and Liquid Phases of Tin from Deep Potentials with {\it ab initio} Accuracy}

\author{Tao Chen\footnote{These authors contributed equally to this work.}}
\affiliation{HEDPS, CAPT, College of Engineering and School of Physics, Peking University, Beijing, 100871, P. R. China}
\author{Fengbo Yuan\footnotemark[1]}
\affiliation{HEDPS, CAPT, College of Engineering and School of Physics, Peking University, Beijing, 100871, P. R. China}

\author{Jianchuan Liu}
\affiliation{HEDPS, CAPT, College of Engineering and School of Physics, Peking University, Beijing, 100871, P. R. China}
\author{Huayun Geng}
\affiliation{National Key Laboratory of Shock Wave and Detonation Physics, Institute of Fluid Physics, CAEP, P.O. Box 919-102, Mianyang 621900, Sichuan, P. R. China}
\author{Linfeng Zhang}
\affiliation{AI for Science Institute, Beijing 100080, P. R. China; DP Technology, Beijing 100080, P. R. China}
\author{Han Wang}
\affiliation{Laboratory of Computational Physics, Institute of Applied Physics and Computational Mathematics, Beijing 100094, P. R. China}
\affiliation{HEDPS, CAPT, College of Engineering and School of Physics, Peking University, Beijing, 100871, P. R. China}
\author{Mohan Chen}
\thanks{Corresponding author. Email: mohanchen@pku.edu.cn}
\affiliation{HEDPS, CAPT, College of Engineering and School of Physics, Peking University, Beijing, 100871, P. R. China}

\date{\today}

\begin{abstract}
Constructing an accurate atomistic model for the high-pressure phases of tin (Sn) is challenging because properties of Sn are sensitive to pressures. 
We develop machine-learning-based deep potentials for Sn with pressures ranging from 0 to 50 GPa and temperatures ranging from 0 to 2000 K.
In particular, we find the deep potential, which is obtained by training the {\it ab initio} data from density functional theory calculations with the state-of-the-art SCAN exchange-correlation functional, is suitable to characterize high-pressure phases of Sn.
We systematically validate several structural and elastic properties of the $\alpha$ (diamond structure), $\beta$, {\it bct}, and $bcc$ structures of Sn, as well as the structural and dynamic properties of liquid Sn.
The thermodynamics integration method is further utilized to compute the free energies of the $\alpha$, $\beta$, {\it bct}, and liquid phases, from which the deep potential successfully predicts the phase diagram of Sn including the existence of the triple-point that qualitatively agrees with the experiment.
\end{abstract}

\maketitle

\section{Introduction}

Tin (Sn) has practical usage in a variety of industrial applications.
For example, Sn has been widely used for soldering joints in electronic devices~\cite{Sn-alloy-solder-1, Sn-alloy-solder-2}.
Moreover, liquid Sn is considered as an alternative liquid-metal plasma-facing component that can be utilized in Tokamak fusion reactors because liquid metals suffer from fewer mechanical failures as compared to solid metals~\cite{14PS-Coenen, 17JCP-Xiaohui, 18JCP-Beatriz, 21JNM-Daye}.
In fact, Sn is an element from group IV of the periodic table and exhibits intriguing properties.
In this group, elements lighter than Sn such as C, Si, and Ge tend to form covalent bonds in a diamond structure with $sp^3$ hybridization orbitals caused by the close energies of valence $s$ and $p$ orbitals.
The tetrahedral bonds of the diamond structure are relatively stronger in the lighter elements than in the heavier elements.
Conversely, the hybrid $sp^3$ orbitals are unlikely to form in the heavier elements since the energies of the $s$ electrons are lower than the $p$ electrons.
As a result, Pb, which is heavier than Sn, prefers to form the face-centered cubic ({\it fcc}) structure~\cite{86B-Pb}.
Sn is located at the borderline between covalent and metallic bonding elements and exhibits unusual bonding properties at different pressures and temperatures.
Therefore, the properties of Sn are sensitive to temperature and pressure conditions~\cite{06B-Pierre}.
In this regard, developing an accurate simulation model for Sn has been a challenge for decades.


The allotropic phase transition of Sn has gained much attention from experiments.
The $\beta$-Sn \MC{phase} is stable at ambient conditions but transforms into the $\alpha$-Sn (diamond structure) with a 27$\%$ volume expansion at temperature below 13.5$^{\circ}$C~\cite{60SSP-Busch}.
When the pressure is larger than 9.8 GPa~\cite{66JAP-Sn-100kbar}, the $\beta$-Sn transforms into a body-centered tetragonal ({\it bct}) phase.
In addition, Sn owns a low melting point of around 505 K~\cite{12JAP-Sn-45GPa}.
The melting temperature increases at higher pressures, implying the existence of a solid-liquid phase boundary.
With the knowledge of the above phases, a triple point is defined \CT{around} 3.02 GPa and 572 K~\cite{14JAP-Sn-xu}.
By increasing the external pressures, body-centered cubic ({\it bcc}) and hexagonal-closest-packed ({\it hcp}) phases were found in some experiments~\cite{11PRB-Sn-157GPa, 13PRB-Sn-138GPa, 16CPB-Sn, 15PRL-Sn-1.5TPa}.
Recently, shock experiments of Sn were investigated in various aspects, such as the solid-liquid phase boundary at high pressure~\cite{19JAP-Lone}, and the static and dynamical phase boundary of the $\beta$-Sn and {\it bct}-Sn phases~\cite{19JSR-Briggs}, etc.

Modeling Sn from first-principles methods is useful in elucidating the underlying mechanisms of several phases in the Sn phase diagram~\cite{22JAC-Guillaume}.
In particular, the density functional theory (DFT)~\cite{64PR-Hohenberg, 65PR-Kohn} \MC{with different exchange-correlation (XC) functionals} has been applied to study solid Sn phases for decades.
For instance, early DFT calculations by using the local density approximations (LDA)~\cite{91PRB-Sn, 93PRB-LDA-Sn, 03PRB-LDA-Sn} or the generalized gradient approximation (GGA)~\cite{08SSC-GGA-Sn} predicted that the phase transition sequence of Sn undergoes $\alpha$ $\rightarrow$ $\beta$ $\rightarrow$ {\it bct} $\rightarrow$ {\it bcc} order at 0 K, which is consistent with the experiments~\cite{66JAP-Sn-100kbar, 13PRB-Sn-138GPa}.
\MC{
Recently, the strongly constrained and appropriately normed (SCAN) functional, which was proposed as a nonempirical and general-purpose XC functional, has been applied to predict the $\beta$-Sn to {\it bct}-Sn transition pressure.~\cite{18B-Perdew}
Impressively, the transition pressure was predicted to be 10.3, 5.4, and 16.2 GPa from the LDA, PBE, and SCAN XC functionals, respectively.
Both LDA and SCAN functionals yield \CT{values close} to the experiment value of 13 GPa~\cite{93PRB-LDA-Sn}, while SCAN yields a closer value to another experimental value around 15-20 GPa~\cite{91B-Sn}. The results point out the importance of using an appropriate XC functional to determine the phase \CT{boundaries} of Sn.
}


Modeling Sn from classical potentials has received wide attentions but still faces challenges~\cite{22JAP-Soulard, 22JAP-Yang}.
For instance, Ravelo and Baskes (RB)~\cite{97L-Sn-meam} proposed a modified embedded-atom method (MEAM) potential for Sn, which is able to describe the phase transition between $\alpha$-Sn and $\beta$-Sn at ambient pressures.
Vella {\it et al.}~\cite{17B-VellaChen-meam} improved the RB potential to accurately simulate liquid Sn but lose some accuracy in describing solid phases of Sn.
Etesami {\it et al.}~\cite{18AM-PbSn-meam} adjusted the RB potential to provide a stable $\beta$-Sn phase near the melting point with better characterized elastic constants in a wide range of temperatures.
Ko {\it et al.}~\cite{18Metals-Sn-meam} developed another MEAM model~\cite{00B-2NNmeam, 01B-2NNmeam-bcc, 10Calphad-2NNmeam}, the parameters of which were optimized via the force-matching method with DFT data.
The new potential improves various properties of pure Sn but is mostly focusing on low-pressure phases.
%
Although the above representative classical potentials are able to simulate solid and liquid phases of Sn, there are still no available force fields that can describe the phase diagram of Sn across a wide range of temperatures and pressures.

Recently, machine-learning-assisted atomistic simulation methods have achieved great success and gained much attention~\cite{11PCCP-Behler,15QC-GAP}.
For example, a recent work adopted the neural network potential to study the thermal conductivity of Sn~\cite{21CMS-Lihong}.
In particular, the deep neural networks are frequently adopted to learn the relations between atomic configurations and the resulting energies, forces, and stresses.
Importantly, by avoiding solving the quantum-mechanics-based equations, these machine learning models exhibit a high efficiency with an {\it ab initio} accuracy.
Among them, the deep potential molecular dynamics (DPMD) method~\cite{18PRL-deepmd, 18CPC-deepmd} stands out as a promising method and has been applied to a variety of systems.
For example, the DPMD method has been applied to study the mechanical properties of the Al-Cu-Mg alloys~\cite{21CPB}, the phase diagram of gallium~\cite{20NC-Niu} and ices~\cite{21L-LinFeng}, the isotope effects in liquid water~\cite{20PRB-Jianhang}, and the structural and dynamic properties of warm dense aluminum~\cite{20JPCM-Qianrui, 21MRE}, etc.
In conclusion, the efficiency and scalability of the DPMD method~\cite{20CPC-Denghui, 20SC-Weile} are much higher than the first-principles methods such as DFT.


In this work, we obtain atomistic models for Sn by using the DPMD method~\cite{18PRL-deepmd, 18CPC-deepmd},
which can describe Sn at temperatures ranging from 0 to 2000 K and pressures ranging from 0 to 50 GPa.
In particular, we systematically test the model trained by DFT data with the usage of the SCAN XC functional, which we refer as the DP-SCAN model for Sn.
We adopt the DP-SCAN model to test various properties of $\alpha$-Sn and $\beta$-Sn solid phases at ambient conditions, as well as the liquid phase at high temperatures.
Besides, the model has been applied to study high-pressure phases of Sn including the {\it bct} and {\it bcc} phases.
Furthermore, the phase diagram of Sn including the $\alpha$, $\beta$, {\it bct} and liquid phases is obtained by using the thermodynamics integration method.
The paper is organized as follows.
In Section~\ref{Methods}, we describe the computational methods utilized in this work.
In Section~\ref{Results}, we discuss the results obtained from the DP model.
Finally, we summarize our results in Section~\ref{Conclusions}.

\section{Computational Methods}
\label{Methods}

\subsection{Density Functional Theory}

We perform DFT calculations for Sn with the Vienna {\it ab initio} simulation package (VASP 5.4.4)~\cite{96B-VASP}.
The projector-augmented wave (PAW) method~\cite{94B-PAW, 99B-USPP} is used to describe the ion-electron interactions with 14 valence electrons for Sn.
We utilize the SCAN meta-generalized gradient approximation (meta-GGA) exchange-correlation functional~\cite{15PRL-SCAN}.
The kinetic energy cutoff is set to 650 eV.
The Gaussian smearing of 0.20 eV is chosen with the smallest allowed spacing between the Monkhorst-pack (MP)~\cite{76B-MP} $k$-points set as 0.10~\AA$^{-1}$ to sample the first Brillouin zone of solid structures.
The total energy is converged to be less than $10^{-6}$ eV \MC{in self-consistent electronic iterations}.

\subsection{Deep Potential Method}
The DP method~\cite{18PRL-deepmd, 18CPC-deepmd} expresses the total energy $E$ of a given system as a sum of atomic contributions, i.e., $E=\sum _i E_i$.
The contribution $E_i$ from atom $i$ depends on an environment matrix $\mathcal{R}_i$, which includes the information of neighboring atoms of atom $i$ within a \MC{cutoff radius}.
The DP model maps $\mathcal{R}_i$ via an embedding net onto a symmetry-preserving descriptor, and then the descriptor is mapped with a fitting network to give $E_i$.
%
Both embedding and fitting neural networks contained three layers with the typical number of neurons being \CT{(25, 50, 100) and (240, 240, 240)}, respectively.
The cut-off radius for each atom was chosen to be 8.0 {\AA}.
The inverse distance $1/r$ decayed smoothly from 2.0 to 8.0 {\AA} in order to remove the discontinuity introduced by the \MC{cutoff}.

A loss function is defined to optimize the embedding and fitting neural networks in the DP method, which takes the form of
\begin{equation}
L(p_\epsilon,p_f,p_\xi)=p_\epsilon\Delta\epsilon^2+\frac{p_f}{3N}\sum_{i}|\Delta\mathbf{F}_i|^2+\frac{p_\xi}{9}||\Delta\xi||^2,
\end{equation}
where $N$ is the number of atoms,
$\epsilon = E/N$ is the energy per atom,
$\mathbf{F}_i$ is the force acting on atom $i$,
$\xi$ is the virial tensor per atom, 
$\Delta$ denotes the difference between the training data and the results predicted by the DP model,
$p_\epsilon$. $p_f$, and $p_\xi$ are tunable prefactors.
\MC{The stochastic gradient descent scheme Adam~\cite{17arX-Adam} was adopted to train the DP model. The final DP model \CT{that was used to evaluate properties of Sn} went through three long training iterations (16, 4, and 4 million steps). In the first and second training iterations, the learning rate was set to 0.001 and 0.0001, respectively; the starting values for $p_\epsilon$. $p_f$, and $p_\xi$ were set to 0.02, 1000, and 0.02, respectively.
During the third training iteration, the starting weights of $p_\epsilon$. $p_f$, and $p_\xi$ were adjusted to 2, 1, and 2, respectively.
}

We adopt an iterative method to generate the DP model, and the initial DP model is trained with a prepared set of DFT data.
In detail, we set up four solid structures of Sn, i.e., an 8-atom $\alpha$-Sn, a 32-atom $\beta$-Sn, a 16-atom {\it bct}-Sn, and a 16-atom {\it bcc}-Sn.
The cell is first expanded or compressed with a scaling factor ranging from 0.82 to 1.02 for the lattice constant $a_0$.
For each of these generated structures, the atomic positions are randomly perturbed by 0.02 {\AA} or the cell vectors are randomly perturbed by 3\% of $a_0$.
Next, a 10-step {\it ab initio} molecular dynamics (AIMD) simulation is performed in the NVT ensemble at 50 K.
Finally, we collect 1468 configurations from all of the AIMD trajectories, which include 548 frames of $\alpha$-Sn, 314 frames of $\beta$-Sn, 270 frames of {\it bct}-Sn, and 336 frames of {\it bcc}-Sn.
In addition, the obtained total energy $E$, virial tensors $\Xi$, and ionic forces $\mathbf{F}_i$ of each atom $i$ together with atomic positions are adopted as the initial training data.

The \MC{concurrent} learning procedure called deep potential generator (DP-GEN)~\cite{19PRM-DPGEN, 20CPC-DPGEN} is applied to perform iterations in order to refine the DP model.
We set up 32 iterations in the DP-GEN workflow.
For each iteration, a set of initial configurations are chosen from the $\alpha$-Sn, $\beta$-Sn, {\it bct}-Sn and {\it bcc}-Sn structures.
Next, five temperatures are chosen for each iteration and the temperatures arise after every 8 iterations.
In total, 20 temperatures ranging from 50 to 1950 K with the interval being 100 K are set.
For each temperature, nine different pressures including 0, 5, 10, 15, 20, 25, 30, 40, and 50 GPa are set.
\MC{Detailed setups of the exploration strategy are listed in Table S1 of Supporting Information (SI)~\cite{23SI}.}
Finally, three steps, i.e., the exploration, labeling, and training processes are performed in each iteration to generate a new DP model.

In the exploration process, we perform DPMD simulations with the LAMMPS package~\cite{95-LAMMPS}.
The isothermal-isobaric ensemble~\cite{84JCP-Nose, 85PRA-Hoover, 94JCP-MTK} is adopted with a time step of 2.0 fs.
In fact, additional DP models are generated in order to yield the model deviation, which depicts the maximum standard deviation of the predicted atomic forces and takes the form of
$\zeta=\mathrm{max}_i\sqrt{\left\langle\left\|\mathbf{f}_i-\bar{\mathbf{f}_i}\right\|^2\right\rangle }$,
where $\bar{\mathbf{f}_i}=\left \langle \mathbf{f}_i \right \rangle$ is the average force acting on the atom $i$ as predicted by different DP models.
By using the model deviation as a criterion, the atomic configurations generated by the above DPMD simulations are divided into three categories, i.e., accurate configurations ($\zeta <$ 0.12 eV/\AA), candidate configurations (0.12 eV/\AA$\ \le \zeta <$ 0.30 eV/\AA), and failed configurations ($\zeta \ge$ 0.30 eV/\AA).
%
%
\MC{Although we start the DP-GEN iterations with solid structures, liquid configurations are also explored in the chosen temperature and pressure ranges. In this regard, it is expected that the resulting DP models should also work for the liquid phase of Sn.}
In the labeling process, 
a selected set of atomic configuration up to 240 are chosen from the candidate configurations and DFT calculations are performed to yield the new training data.
In the training step, 
$4\times10^5$ steps are performed to train the new DP models.
%
%
The final DP-SCAN model is generated by using 6647 DFT configurations, which include 2212, 1530, 1531, and 1374 frames from the $\alpha$-Sn, $\beta$-Sn, {\it bct}-Sn, and {\it bcc}-Sn structures, respectively.
\MC{We also generated a DP-PBE model based on the PBE exchange-correlation functional~\cite{96L-PBE}, and the results can be found in SI~\cite{23SI}.}

%
%
\section{Results and Discussion}
\label{Results}
\subsection{Solid Sn}

\begin{table*}[h!t!p]
\centering
\caption{Properties of the $\alpha$-Sn, $\beta$-Sn, {\it bct}-Sn, and {\it bcc}-Sn structures including the cohesive energy $E_c$ (in eV), the lattice constants $a$ and $c$ (in \AA) and the $c/a$ ratio, the equilibrium volume $V_0$ (in \AA$^3$), the bulk modulus $B_0$ (in GPa), and the elastic constants $C_{11}$, $C_{12}$, $C_{13}$, $C_{33}$, $C_{44}$, and $C_{66}$ (in GPa) as computed by the DFT with the SCAN functional, the deep potential model (DP-SCAN), four empirical force fields (RB~\cite{97L-Sn-meam}, Vella~\cite{17B-VellaChen-meam}, Etesami~\cite{18AM-PbSn-meam}, and Ko~\cite{18Metals-Sn-meam}), as well as the available experimental data (Exp.).}
\setlength{\tabcolsep}{6pt}
\renewcommand\arraystretch{1.1}
\begin{tabular}{ccccccccc}
\hline
$\alpha$-Sn & DFT & DP-SCAN & RB~\cite{97L-Sn-meam} & Vella~\cite{17B-VellaChen-meam} & Etesami~\cite{18AM-PbSn-meam} & Ko~\cite{18Metals-Sn-meam} & Exp.\\
\hline
$E_c$ & -3.460 & -3.464 & -3.140 & -3.219 & -3.209 & -3.135 & -3.140~\cite{Sn-exp-2}\\
$a$    &  6.566 & 6.560 &  6.483 &  6.304 &  6.430 &  6.581 & 6.483~\cite{Sn-exp-1} \\
$V_0$  & 35.384 & 35.285 & 34.059 & 31.315 & 33.231 & 35.628 & 34.059~\cite{Sn-exp-1} \\
$B_0$  & 38.2  & 42.6 & 42.2  & 44.2  & 43.6  & 40.6  & 42.6~\cite{Sn-exp-8} \\
$C_{11}$  & 52.7  & 63.2 & 70.4  & 64.9  & 82.0  & 50.4  & 69.1~\cite{Sn-exp-8} \\
$C_{12}$  & 31.4  & 32.3 & 28.2  & 33.9  & 24.5  & 35.7  & 21.3~\cite{Sn-exp-8} \\
$C_{44}$  & 22.5  & 29.6 & 36.7  & 42.5  & 94.8  & 10.5  & 42.6~\cite{Sn-exp-8}\\
\hline
$\beta$-Sn & DFT & DP-SCAN & RB~\cite{97L-Sn-meam} & Vella~\cite{17B-VellaChen-meam} & Etesami~\cite{18AM-PbSn-meam} & Ko~\cite{18Metals-Sn-meam} & Exp.\\
\hline
$E_c$  & -3.385 & -3.397 & -3.085 & -3.115 & -3.091 & -3.102 & -3.10~\cite{Sn-exp-4},-3.130~\cite{03PRB-LDA-Sn} \\
$a$    &  5.909 & 5.895  &  5.920 &  5.682 &  5.914 &  5.859 & 5.831~\cite{Sn-exp-1},5.8119~\cite{Sn-exp-3} \\
$c$    &  3.164 & 3.170  &  3.235 &  3.334 &  3.237 &  3.206 & 3.184~\cite{Sn-exp-1},3.1559~\cite{Sn-exp-3} \\
$c/a$  &  0.536 & 0.538  &  0.546 &  0.587 &  0.547 &  0.547 & 0.546~\cite{Sn-exp-1},0.543~\cite{Sn-exp-3} \\
$V_0$  & 27.619 & 27.543 & 28.345 & 26.911 & 28.301 & 27.514 & 27.064~\cite{Sn-exp-1},26.650~\cite{Sn-exp-3} \\
$B_0$  & 53.1  & 52.5  & 64.6  & 65.6  & 64.8  & 57.1  & 57.0~\cite{Sn-exp-8},57.037~\cite{Sn-exp-7} \\
$C_{11}$  & 108.6  & 110.6  & 109.0  & 123.2  & 132.6  & 89.8  & 73.2~\cite{Sn-exp-8} \\
$C_{12}$  & 11.5  & 19.2  & 62.7  & 34.4  & 47.1  & 46.7  & 59.8~\cite{Sn-exp-8} \\
$C_{13}$  & 33.6  & 27.6   & 24.5  & 54.1  & 16.7  & 36.9  & 39.1~\cite{Sn-exp-8} \\
$C_{33}$  & 103.4  & 101.8 & 139.8 & 59.2  & 157.3 & 93.8  & 90.6~\cite{Sn-exp-8} \\
$C_{44}$  & 24.4  & 23.1  & 0.8  &  4.1  &  1.2  &  7.8  & 21.9~\cite{Sn-exp-8} \\
$C_{66}$  & 24.7  & 28.5  & 22.3  & 21.3  & 27.8  & 10.6  & 23.8~\cite{Sn-exp-8} \\
\hline
{\it bct}-Sn & DFT & DP-SCAN & RB~\cite{97L-Sn-meam} & Vella~\cite{17B-VellaChen-meam} & Etesami~\cite{18AM-PbSn-meam} & Ko~\cite{18Metals-Sn-meam} & Exp.\\
\hline
$E_c$  & -3.278  & -3.290  & -3.080 & -3.084 & -3.080 & -3.082 & - \\
$a$    &  4.016  & 4.008   &  3.433 &  3.055 &  3.439 &  3.276 & 3.70~\cite{66JAP-Sn-100kbar} \\
$c$    &  3.361  & 3.365   & 4.870  &  5.924 &  4.853  &  4.634  & 3.37~\cite{66JAP-Sn-100kbar} \\
$c/a$  &  0.837  & 0.840   & 1.419  &  1.939 &  1.411  &  1.414  & 0.911~\cite{66JAP-Sn-100kbar} \\
$V_0$  & 27.104  & 27.027  & 28.698 & 27.645 & 28.698  & 24.872  & 23.068~\cite{66JAP-Sn-100kbar},26.25~\cite{13PRB-Sn-138GPa} \\
$B_0$  &  53.2 & 46.2 & 73.5  & 70.1  & 73.5  & 68.0  & 63~\cite{13PRB-Sn-138GPa} \\
$C_{11}$  & 61.4  & 57.3  & 99.1  & 125.7  & 103.9  & 172.5 & - \\
$C_{12}$  & 37.3  & 34.9  & 48.0  & 35.6  & 49.3  & -19.8  & - \\
$C_{13}$  & 49.1  & 38.5  & 73.5  & 59.0  & 73.5  & 51.2  & - \\
$C_{33}$  & 84.6  & 74.3  & 73.6  & 72.3  & 73.5  & 101.9  & - \\
$C_{44}$  & 11.3  & 11.3  & 24.6  &  0.5  & 30.8  & 95.8  & - \\
$C_{66}$  & 26.9  & 23.7  & 0.03  & 12.2  & 0.03  & 25.5  & - \\
\hline
{\it bcc}-Sn & DFT & DP-SCAN & RB~\cite{97L-Sn-meam} & Vella~\cite{17B-VellaChen-meam} & Etesami~\cite{18AM-PbSn-meam} & Ko~\cite{18Metals-Sn-meam} & Exp.\\
\hline
$E_c$   & -3.265 & -3.274 & -3.080 & -3.075 & -3.080 & -3.082 & - \\
$a$     &  3.774 &  3.780 &  3.847 &  3.791 &  3.847 &  3.752 & 3.659~\cite{13PRB-Sn-138GPa} \\
$V_0$   & 26.879 & 27.002 & 28.458 & 27.252 & 28.468 & 26.415 & 24.5~\cite{13PRB-Sn-138GPa} \\
$B_0$   & 47.9   & 46.5  & 73.9  & 74.9  & 73.9  & 63.2  & 92~\cite{13PRB-Sn-138GPa} \\
$C_{11}$   & 42.3  & 34.7 & 81.9  & 90.4  & 83.8  & 72.9  & - \\
$C_{12}$   & 50.3  & 52.4 & 69.9  & 67.1  & 69.0  & 58.4  & - \\
$C_{44}$   & 28.6  & 7.1  & 29.7  &  6.0  & 36.5  & 26.4  & - \\
\hline
\end{tabular}
\label{CrystalTable}
\end{table*}

\begin{figure}[h!t]
  \begin{center}
    \includegraphics[width=8cm]{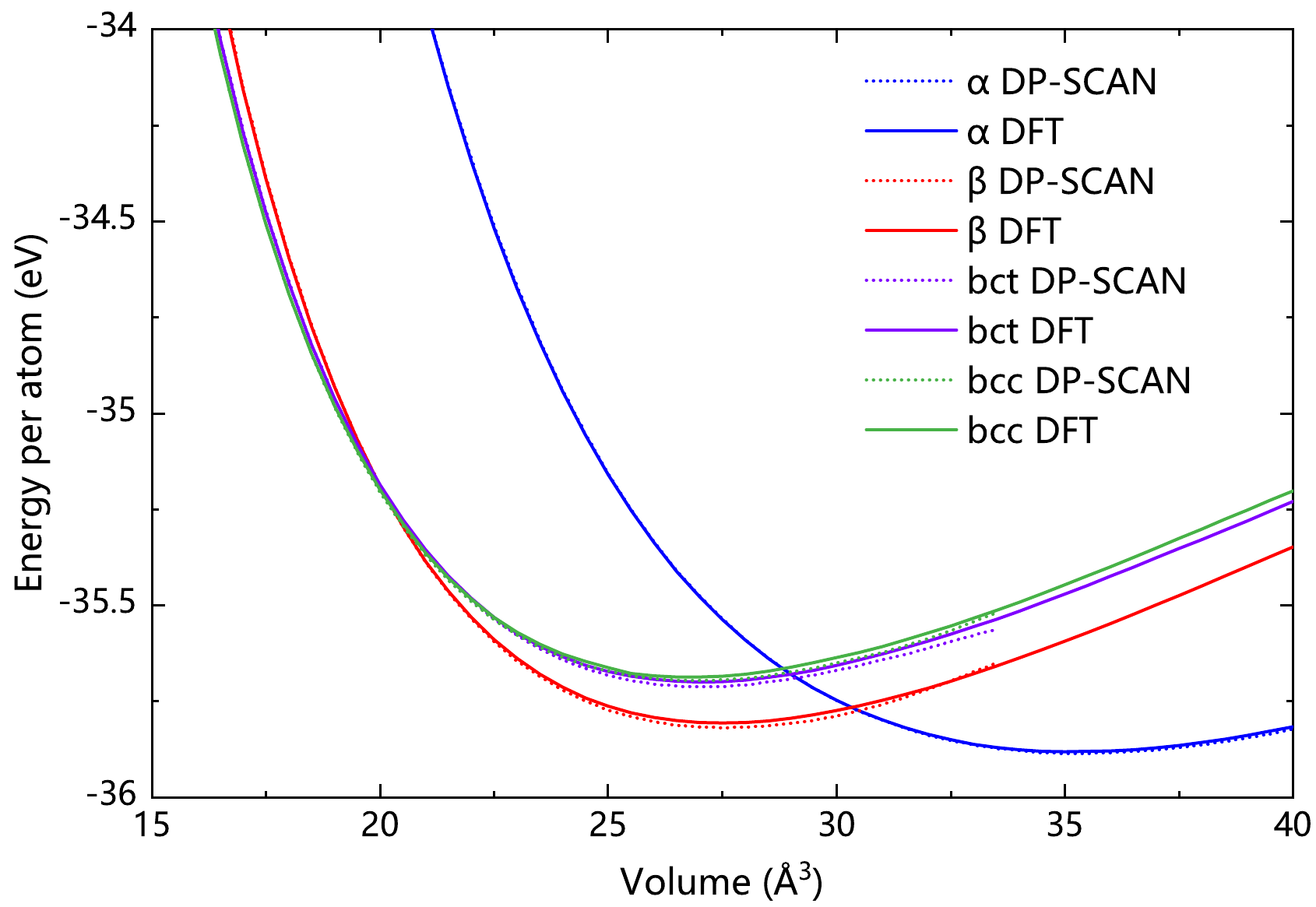}
  \end{center}
\caption{(Color online) \MC{Equation of state} of the $\alpha$-Sn, $\beta$-Sn, {\it bct}, and {\it bcc} phases as computed by the DP-SCAN model and the DFT method with the SCAN XC functional.}
\label{fig:deepmd_eos}
\end{figure}

Fig.~\ref{fig:deepmd_eos} illustrates the \MC{EOS (Equation of State)} of four crystal structures of Sn as calculated by using the DP-SCAN model and the DFT method with the SCAN functional.
In general, the DP-SCAN model well reproduces the DFT results for \CT{all four} crystalline structures considered here, \CT{namely} the $\alpha$-Sn, $\beta$-Sn, {\it bct}-Sn, and {\it bcc}-Sn structures.
To be specific, both methods predict the $\alpha$ phase to be the most stable phase, \CT{followed} by $\beta$-Sn, {\it bct}-Sn and {\it bcc}-Sn structures.
The DP-SCAN model reproduces the small energy differences between the {\it bct}-Sn and {\it bcc}-Sn.
\CT{Since only a few training data were prepared for Sn at larger volumes, we do not expect the DP-SCAN potential to perform well for the crystal structures at larger volumes.}

Several bulk properties of solid Sn phases are listed in Table~\ref{CrystalTable}, which includes the cohesive energy $E_c$, the lattice constants ($a$, $c$, and the $c/a$ ratio), the equilibrium volume $V_0$, the bulk modulus $B_0$, and the elastic constants of the $\alpha$-Sn, $\beta$-Sn, {\it bct}-Sn and {\it bcc}-Sn structures.
The data were obtained from the DP-SCAN model, the DFT method with the SCAN XC functional, four empirical force fields (RB~\cite{97L-Sn-meam}, Vella~\cite{17B-VellaChen-meam}, Etesami~\cite{18AM-PbSn-meam}, and Ko~\cite{18Metals-Sn-meam}) as well as the available experimental data.
\MC{We also calculate the bulk properties of solid Sn phases from the PBE functional and the results are listed in Table S2 of SI~\cite{23SI}.}

The cohesive energy is defined as the difference of energy per atom in a bulk phase and the energy of a single atom.
In DFT calculations, the spin-polarized calculations were performed for a single Sn atom to yield the total energy.
Both DFT and DP-SCAN methods give rise to \CT{values close} of -3.460 and -3.464 eV for the $\alpha$-Sn, respectively.
The DFT method with the SCAN functional does not rely on empirical parameters and the deep neural network well reproduces the DFT result.
Although the two cohesive energies are predicted to be about 0.3 eV lower than the experimental value of -3.140 eV~\cite{Sn-exp-2}, we note that the results follow the same trend as explained by previous DFT calculations~\cite{03PRB-LDA-Sn,17B-VellaChen-meam}.
%
%
%

The energy orderings of different solid phases can be analyzed by comparing their cohesive energies.
On the one hand, the energy orderings of the $\alpha$-Sn, $\beta$-Sn, {\it bct}-Sn and {\it bcc}-Sn phases obtained by the DP-SCAN model match well with the DFT results.
Next, the relative energies between the $\alpha$-Sn and the other three solid phases including the $\beta$-Sn, {\it bct}-Sn, and {\it bcc}-Sn phases are 0.075 (0.067), 0.182 (0.174), and 0.195 (0.190) eV as obtained from the DFT (DP-SCAN) method.
On the other hand, \CT{all four} empirical force fields listed in Table~\ref{CrystalTable} yield correct energy ordering between the $\alpha$ and $\beta$ phases of Sn, but the results deviate after including the {\it bct} and {\it bcc} phases.
For example, the RB~\cite{97L-Sn-meam}, Etesami~\cite{18AM-PbSn-meam}, and Ko~\cite{18Metals-Sn-meam} force fields all lead to the same cohesive energy for the {\it bct}-Sn and {\it bcc}-Sn structures, which does not agree with the DFT data while experimental data are not available.

For the cell volumes of the $\alpha$-Sn and $\beta$-Sn structures, the RB potential yields the same volume of 34.059~\AA$^3$ as compared to the experimental value~\cite{Sn-exp-1}, while the DP-SCAN model yields a slightly larger value of 35.285~\AA$^3$.
However, the DP-SCAN model gives a more accurate cell volume of 27.543~\AA$^3$ for the $\beta$-Sn, which is closer to the experimental values of 27.064~\cite{Sn-exp-1} or 26.650~\AA$^3$~\cite{Sn-exp-3} as compared to the value of 28.345~\AA$^3$ from the RB potential.
In fact, we find both DP-SCAN and empirical force fields reproduce quite well the cell volume, the lattice constants, and the bulk modulus of the $\alpha$-Sn and $\beta$-Sn structures as compared to the experiment.
More details can be found in Table~\ref{CrystalTable}.

A significant defect of these empirical force fields is the lack of enough accuracy in describing the lattice constants ($a$ and $c$) of the {\it bct} phase of Sn.
A typical example is the $c/a$ ratio ($a$ and $c$ are lattice constants of {\it bct}-Sn).
The predicted $c/a$ ratio ranges from 1.411 to 1.939 from the four empirical force fields, while the result from the experiment is $c/a = 0.91$~\cite{66JAP-Sn-100kbar}.
Notably, the $c/a$ values predicted by the DP-SCAN model and the DFT method are 0.840 and 0.837, respectively.
In addition, the DP-SCAN predicts the lattice constant $c$ of {\it bct}-Sn to be 3.365~\AA, which is close to the experimental value of 3.37~\AA~\cite{66JAP-Sn-100kbar}; the lattice constant $a$ is predicted to be 4.008~\AA, larger than the experimental value of 3.70~\AA~\cite{66JAP-Sn-100kbar}.
Nevertheless, the data from the DP-SCAN model are much closer \CT{to} the experimental value than those from empirical force fields.
This is expected because the lattice constants of the {\it bct} phase were not adopted to tune the parameters in these empirical force fields while the DFT method \CT{owes its predictive power to} quantum mechanics calculations without empirical parameters.

In terms of the elastic constants, we have the following findings. 
First, we observe all of the models yield reasonable values for the $C_{11}$ and $C_{12}$ of the $\alpha$ phase, but large deviations are found for the $C_{44}$. 
For instance, while the experimental value is 42.6 GPa~\cite{Sn-exp-8}, the Etesami and Ko models predict the $C_{44}$ to be 94.8 and 10.5 GPa, respectively. 
Meanwhile, the values from DFT and DP-SCAN are 22.5 and 29.6 GPa, respectively.
Second, for the $\beta$ phase, \CT{all four} empirical potentials yield a small value of $C_{44}$ ranging from 0.8 to 7.8 (in GPa).
Notably, the $C_{44}$ value from the DP-SCAN model is 23.1 GPa, which is close to the experimental value of 21.9 GPa~\cite{Sn-exp-8}. 
\CT{Third, we notice that the DP-SCAN model predicts a substantially smaller value of $C_{44}$ (7.1 GPa) for $bcc$-Sn when compared to the value of 28.6 GPa from SCAN.}
Furthermore, all of the models yield a larger value for the $C_{11}$ while the DFT and DP-SCAN models substantially underestimate the $C_{12}$ value. 
In terms of the DFT accuracy, the reason for the above deviations may be caused by the orbital localization error of density functionals~\cite{20JCP-Geng}.
Other elastic constants of $\beta$-Sn include $C_{13}$, $C_{33}$, and $C_{66}$ are well described by the DFT and DP-SCAN models.

\begin{figure}[h!t]
  \begin{center}
    \includegraphics[width=8cm]{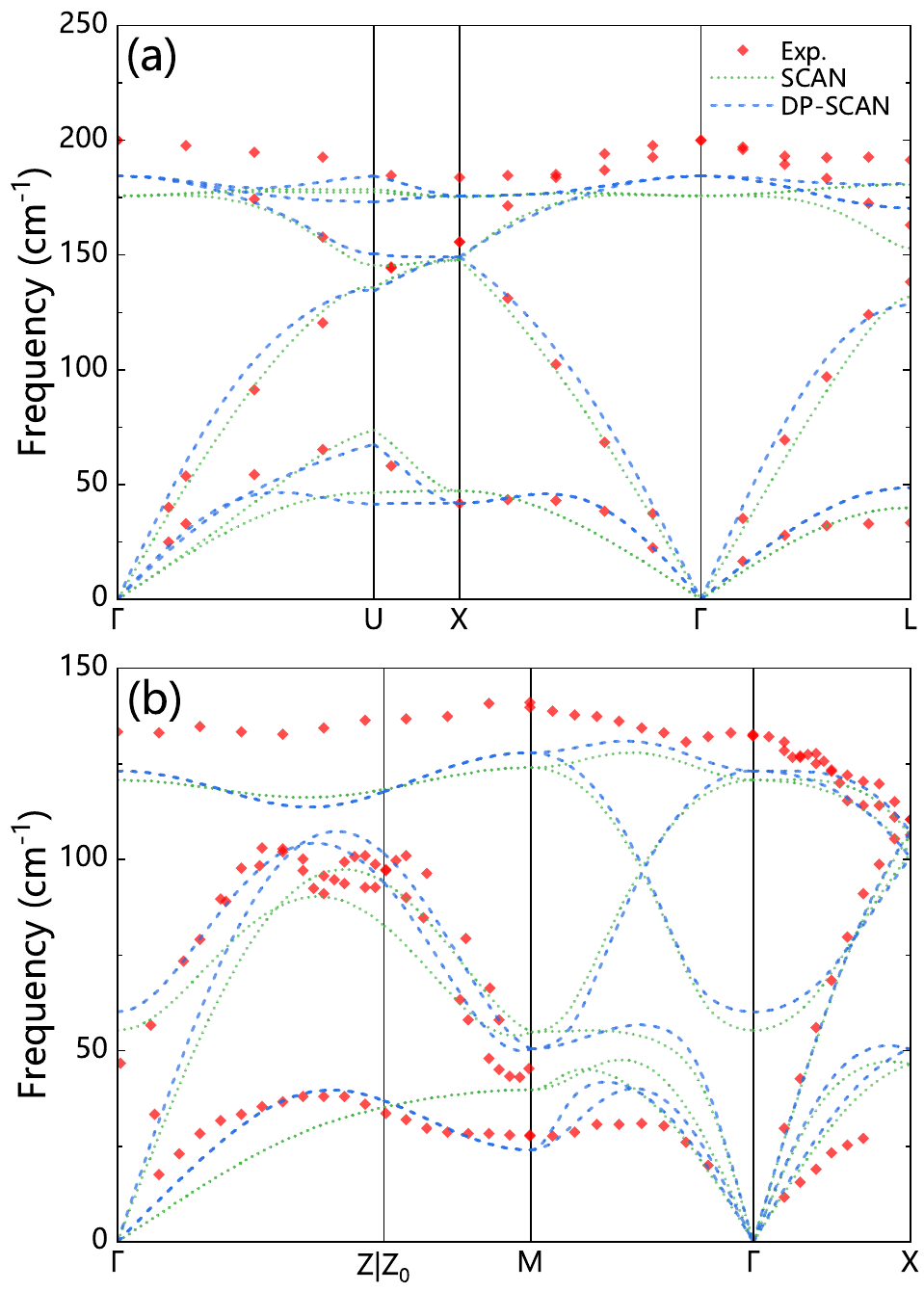}
  \end{center}
\caption{(Color online) \MC{Phonon dispersion of (a) $\alpha$-Sn and (b) $\beta$-Sn from DFT calculations with the SCAN functional (green dotted lines) and the DP-SCAN model (blue dashed lines). 
The experimental data for the $\alpha$-Sn and $\beta$-Sn phases (red dots) are from Refs.~\cite{71B-alpha_Sn} and \cite{67PR-beta_Sn}, respectively.}
\MC{The coordinates of special $k$-points are listed in Tables S3 and S4 of SI~\cite{23SI}.}}
\label{fig:phonon}
\end{figure}

\MC{The phonon dispersions of the $\alpha$-Sn and $\beta$-Sn phases as computed from both DFT and the DP-SCAN model, as well as the experimental data~\cite{71B-alpha_Sn, 67PR-beta_Sn} are shown in Fig.~\ref{fig:phonon}. 
We generate a $2\times2\times2$ supercell and use the Phonopy package (v.2.15.1)~\cite{15SM-phonopy} to compute the phonon dispersion relations.
For the $\alpha$-Sn ($\beta$-Sn) structure, we use a $5\times5\times5$ ($7\times7\times7$) $\Gamma$-centered $k$-point sampling and the kinetic energy cutoff is set to 650 (1200) eV.
We find the phonon dispersion obtained from the DP-SCAN model are close to the DFT SCAN results for the $\alpha$-Sn. 
In addition, the results from both models match well with the experimental data. 
For the $\beta$-Sn phase, We notice that the SCAN and DP-SCAN models exhibit some deviations in characterizing the phonon dispersions of $\beta$-Sn from the $\Gamma$ point to the M point; however, since the machine-learned DP-SCAN model is generated in a wide range of temperatures and pressures, we consider the results are reasonable. 
When compared to the experimental data, the calculated frequencies from both models in the high-frequency region are lower than the experimental ones, but the remaining features in the low-frequency region match reasonably with available experimental data.
}

\subsection{Liquid Sn}

We performed NPT MD simulations with the DP-SCAN model using the LAMMPS package for liquid Sn at temperatures ranging from 573 to 1873 K and pressures of 0, 10, and 20 GPa.
The Nos\'e-Hoover thermostat and \MC{MTK} barostat~\cite{84JCP-Nose, 85PRA-Hoover, 94JCP-MTK} were used.
For a given temperature and a fixed pressure, the MD trajectory length is 250 ps with a time step of 0.5 fs. 
\CT{Although a time step of 0.5 fs was used in this study, we found a time step of 2 fs also yielded reasonable results (see Fig. S2 of SI~\cite{23SI}).}
We took $4\times10^4$ configurations from the last 200 ps of the simulations to compute properties of liquid Sn.

The structure of liquid Sn can been seen by the radial distribution function
\begin{equation}
g(r)=\frac{V}{4\pi r^2N^2}\bigg\langle\sum_{i=1}^N
\sum_{j=1,j\neq i}^{N}\delta(r-|\mathbf{r}_i-\mathbf{r}_j|)\bigg\rangle,
\end{equation}
where $V$ is the cell volume, $N$ is the number of atoms,
$\mathbf{r}_i$ and $\mathbf{r}_j$ are atomic coordinates of atoms $i$ and $j$,
and $\langle\cdots\rangle$ means the ensemble average.
We performed simulations on a 432-atom system to compute the radial distribution functions and liquid densities; we also performed simulations for a 250-atom system and no significant size effects were observed.

\begin{figure}[h!t]
    \begin{center}
	    \includegraphics[width=8cm]{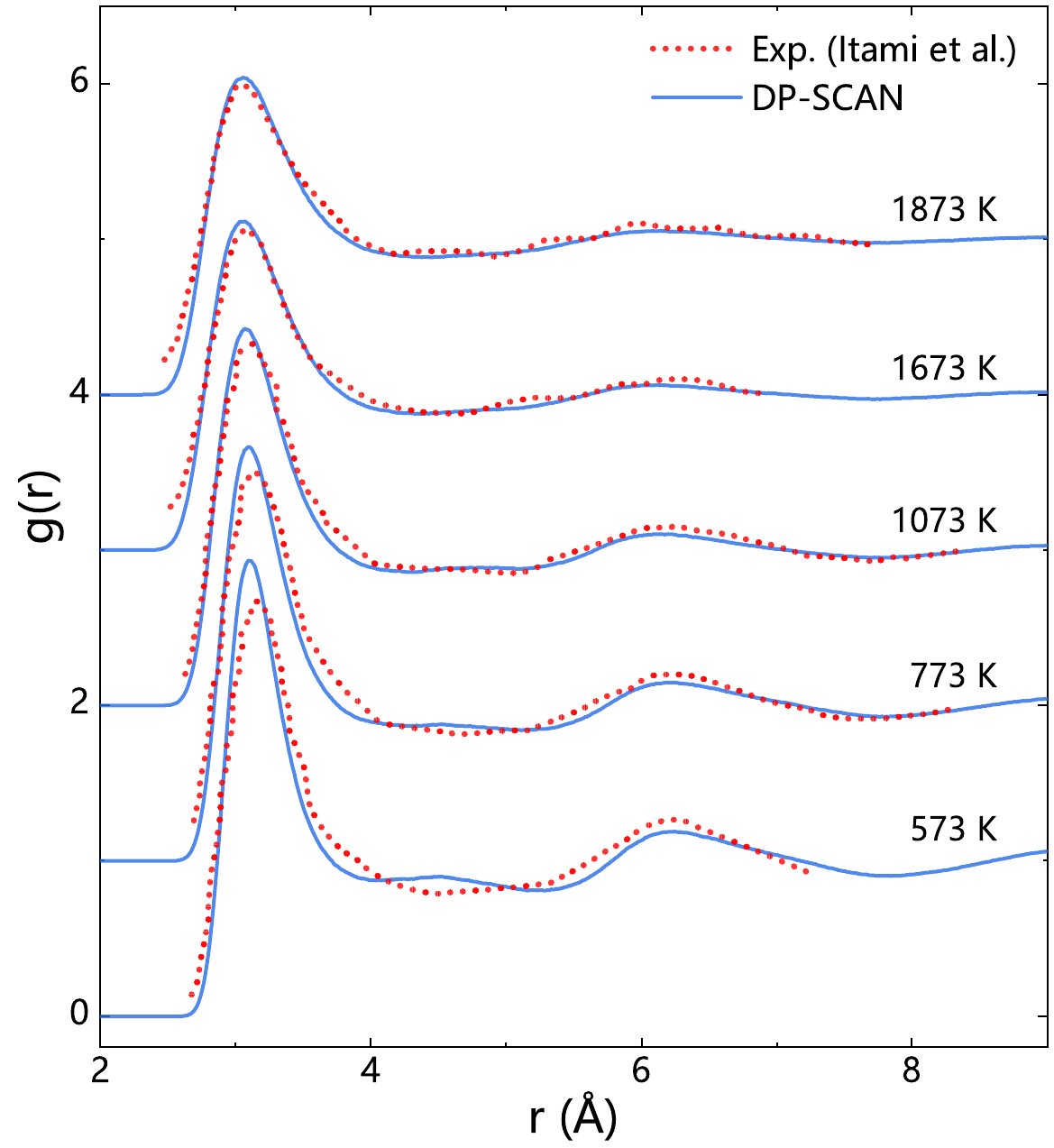}
	\end{center}
	\caption{(Color online) Radial distribution functions $g(r)$ of 432-atom liquid Sn systems as obtained from molecular simulations using the DP-SCAN model (blue solid lines). 
	The experimental data~\cite{03PRB-Itami} (red dotted lines) are shown for comparison. 
	The temperatures are set to 573, 773, 1073, 1673, and 1873 K. 
	The pressure is set to zero.
	}\label{fig:pdf}
\end{figure}

\begin{figure}[h!t]
    \begin{center}
	    \includegraphics[width=8cm]{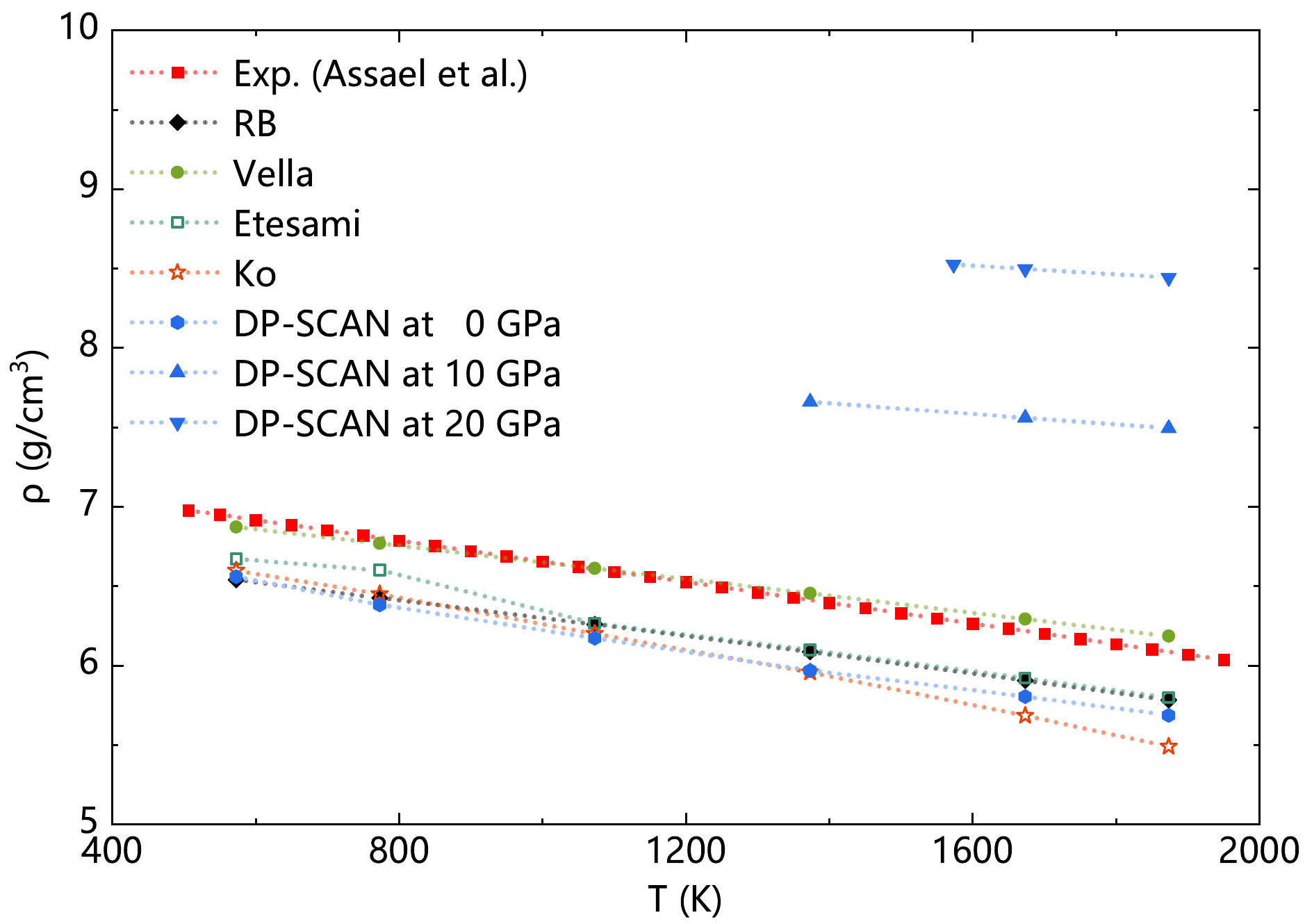}
	\end{center}
	\caption{
(Color Online) Comparison of the predicted liquid densities from the DP-SCAN model, the RB~\cite{97L-Sn-meam}\CT{, Vella~\cite{17B-VellaChen-meam}, Etesami~\cite{18AM-PbSn-meam}, and Ko~\cite{18Metals-Sn-meam}
empirical force fields}, and the experimental data~\cite{10JPC-Assael}. 
Three different pressures (0, 10, and 20 GPa) are adopted in simulations using the DP-SCAN model.
	}\label{fig:density}
\end{figure}

The radial distribution functions of the DP model at 573, 773, 1073, 1673, and 1873 K and the experimental data~\cite{03PRB-Itami} are shown in Fig.~\ref{fig:pdf}.
The pressure was set to zero.
At 573 and 773 K, the DP model slightly overestimates the height of the first peak. 
In addition, a bump structure appears \CT{around} 4~\AA~from experiments and the DP-SCAN model predicts a bump structure \CT{around} 4.5~\AA, indicating that the DP-SCAN model still shows some deviations with the experimental data~\cite{03PRB-Itami}. 
However, the DP-SCAN model yields a similar peak \CT{around} 6.2~\AA, which agrees well with the experiments. 
Notably, at higher temperatures of 1073, 1673, and 1873 K, the peak heights from DP-SCAN match perfectly with the experimental data. 
In conclusion, we consider the DP-SCAN model gives reasonable local structures of liquid Sn at zero pressure. 
We compare the densities of liquid Sn from the DP-SCAN model to the experimental data~\cite{10JPC-Assael} and the \CT{RB~\cite{97L-Sn-meam}, Vella~\cite{17B-VellaChen-meam}, Etesami~\cite{18AM-PbSn-meam}, and Ko~\cite{18Metals-Sn-meam} force fields in Fig.~\ref{fig:density}.
We find only the Vella model reproduces the experimental data well, other models underestimate the density of Sn at 0 GPa by around $6\%$ as compared to the experimental data~\cite{10JPC-Assael}.}
In addition, we use the DP-SCAN model to predict the liquid Sn densities at 10 and 20 GPa, which shows higher densities at higher pressures.

\begin{figure}[h!t]
    \begin{center}
	    \includegraphics[width=8cm]{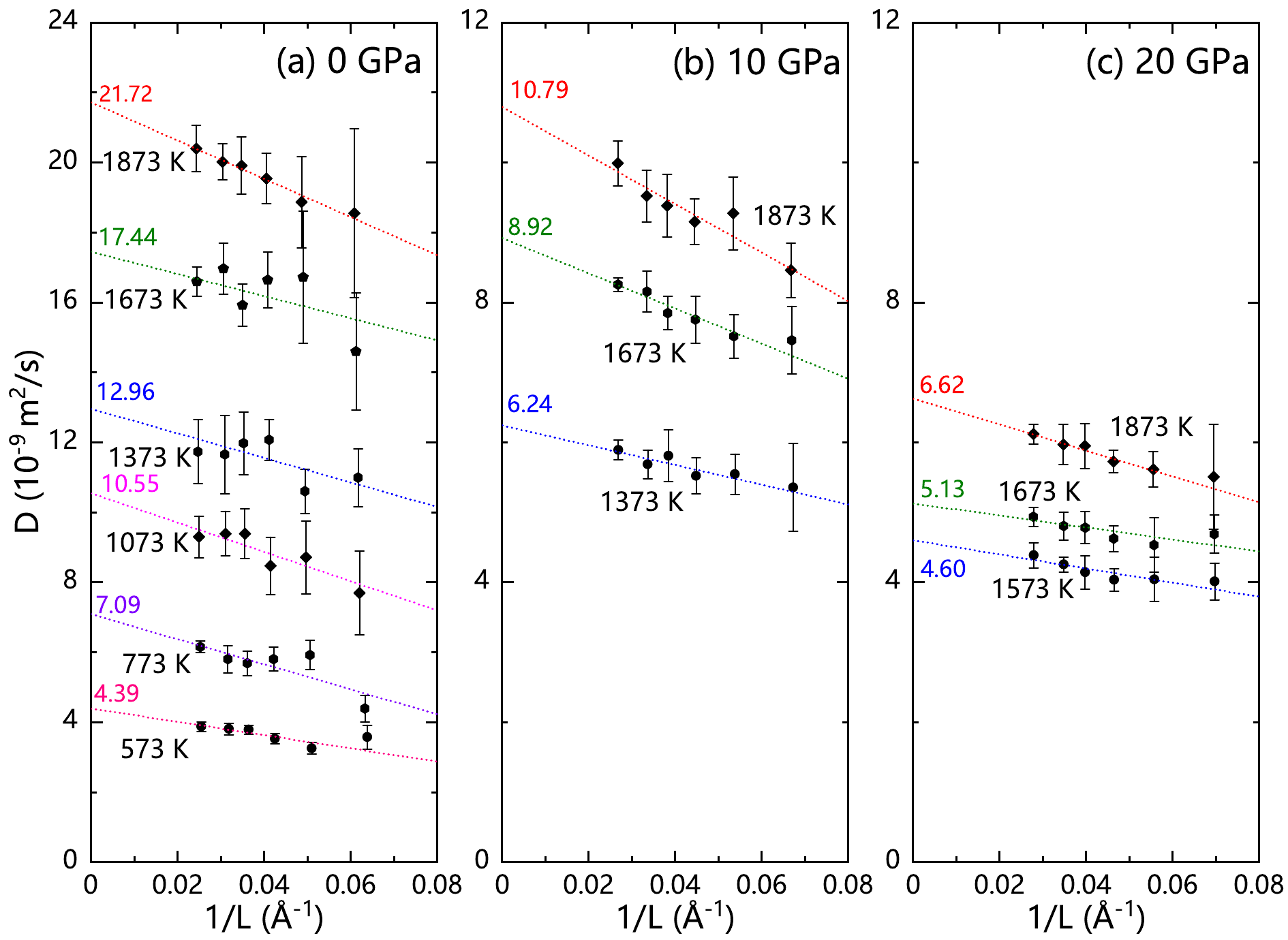}
	\end{center}
	\caption{
(Color online) Self-diffusion coefficients $D$ of liquid Sn calculated from different system sizes (128-2000 atoms) and at different pressures, including (a) 0 GPa, (b) 10 GPa, and (c) 20 GPa. 
The temperature ranges from 573 to 1873 K.
$L$ is the cell length of the cubic cell. 
The extrapolation data of diffusion coefficients are listed as $1/L$ approaches zero.	
    }\label{fig:diff_cal}
\end{figure}

In terms of dynamic properties, we calculated the self-diffusion coefficients of liquid Sn ($D^{calc}$) with temperature and pressure conditions as used in computing the liquid Sn densities shown in Fig.~\ref{fig:density}. 
The self-diffusion coefficients were obtained through the mean square displacements of atoms with the formula~\cite{1905AP-Einstein} written as
\begin{equation}
D^{calc}=\frac{\langle\Delta\mathbf{r} ^2\rangle}{6t},
\end{equation}
where $\Delta\mathbf{r}$ is the displacement \CT{of the \Mark{individual} atoms at time $t$\Mark{~\cite{01B-Van, 17PRM-Marcolongo}}}.
\Mark{In detail, the MD trajectory length was 450 ps with a time step of 1 fs. 
In practice, we divided the last 400 ps trajectory into 8 segments, each with a trajectory length of 50 ps.}
\CT{We then computed $D$ through a linear fitting for the slope of the MSD for each segment and took the average.}
To reduce the \CT{finite} size effects, we adopted the formula in Ref.~\cite{04JPCB-Yeh}:
\begin{equation}
	D^{corr}=D^{calc}+\frac{k_BT\xi}{6\pi\eta L},
\end{equation}
where $D^{corr}$ is the corrected self-diffusion coefficient without artificial size effects, $D^{calc}$ is the self-diffusion coefficient calculated from a specific system, $k_B$ is the Boltzmann constant, $T$ is the temperature, $\xi$ is a constant equal to 2.837297, $\eta$ is the shear viscosity, and $L$ is the length of the cubic cell.
For each temperature and pressure, at least 6 system sizes with the number of atoms ranging from 128 to 2000 were used in order to eliminate the size effects in the diffusion coefficient, the results are illustrated in Fig.~\ref{fig:diff_cal}.
We observe a linear relationship between the computed diffusion coefficients $D^{calc}$ and the $1/L$ parameter. 
By extrapolating $D^{calc}$ to the infinite system size ($1/L=0$), we obtain the corrected self-diffusion coefficient 
$D^{corr}$ which are shown in Fig.~\ref{fig:diff_cal}.

\begin{figure}[h!t]
    \begin{center}
	    \includegraphics[width=8cm]{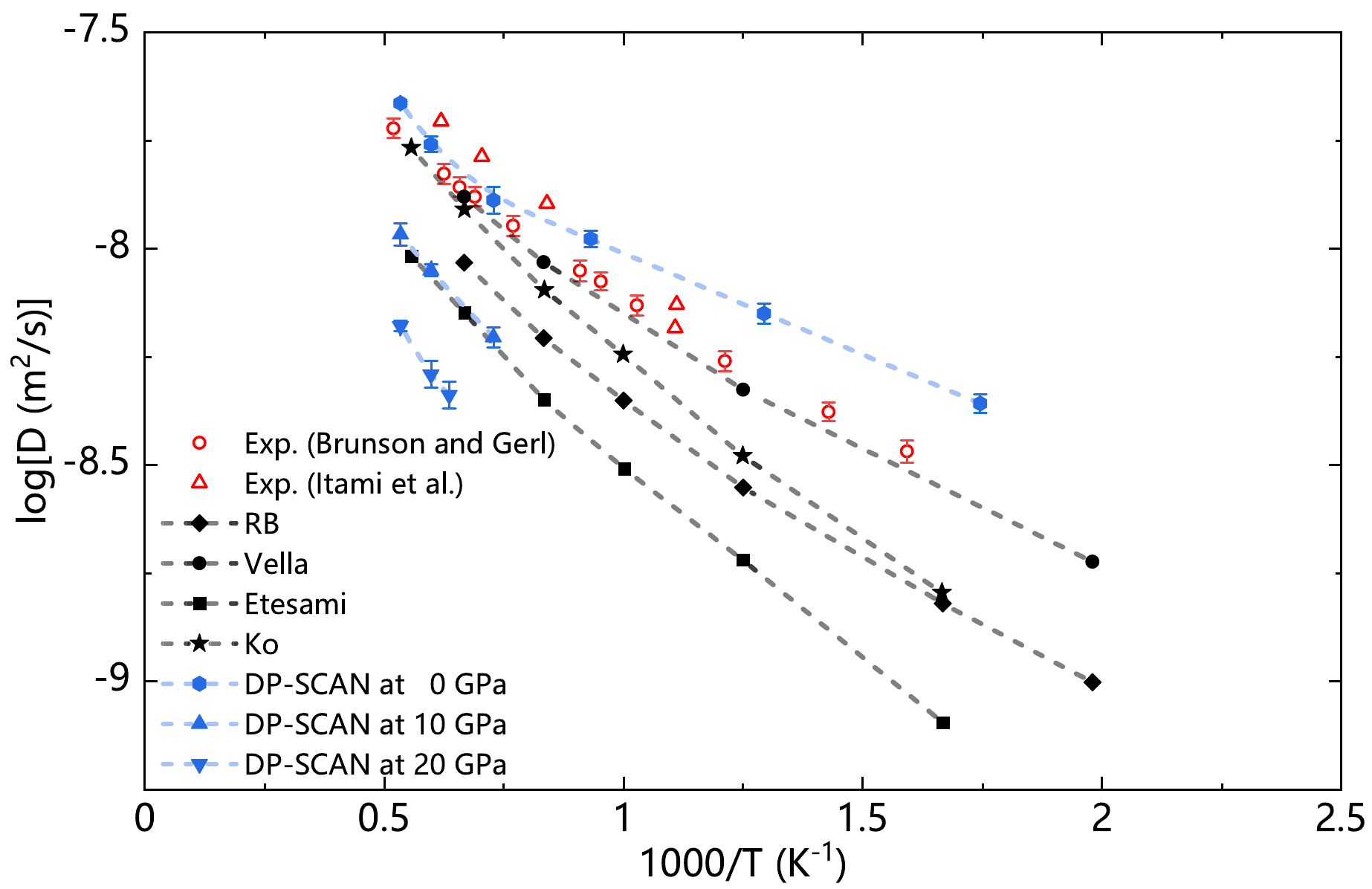}
	\end{center}
	\caption{
(Color online) Comparison of the predicted self-diffusion coefficients of Sn from the DP-SCAN model and the empirical force fields (\CT{RB, Vella, Etesami, and Ko}), as well as the experimental data 
(Bruson and Gerl~\cite{80PRB-Bruson}, and Itami {\it et al.}~\cite{98JJSMA-onishi}).
	}\label{fig:Diff}
\end{figure}

\CT{Fig.~\ref{fig:Diff} illustrates the predicted self-diffusion coefficients from the DP-SCAN model and the four classical potentials (RB, Vella, Etesami, and Ko), as well as the available experimental data~\cite{80PRB-Bruson, 98JJSMA-onishi}.
In general, the self-diffusion coefficients of Sn at 0 GPa from the DP model and the Vella force field are consistent with the experimental data, while the other three empirical force fields underestimate the self-diffusion coefficients especially at relatively lower temperatures.
Moreover, the DP model predicts that the self-diffusion coefficient of Sn decreases at higher pressures (10 and 20 GPa).
}

\subsection{Phase Diagram of Sn}

\begin{figure}[h!t]
  \begin{center}
    \includegraphics[width=8cm]{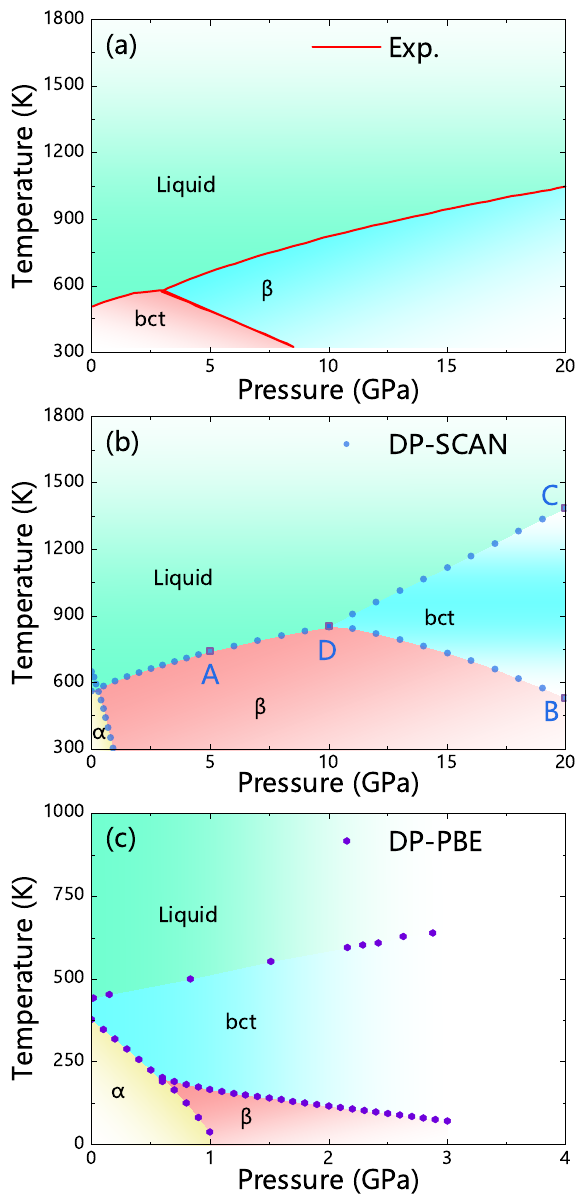}
  \end{center}
\caption{\CT{(Color online) Phase diagram of Sn with temperatures ranging from 0 to 2000 K and pressures ranging from 0 to 20 GPa from (a) experiments~\cite{14JAP-Sn-xu} and (b) the DP-SCAN model. Four phases of Sn are presented including the $\alpha$-Sn, $\beta$-Sn, {\it bct}-Sn, and liquid phases. The coexistence points on the phase diagram of (b) are labeled as A (5 GPa, 741 K), B (20 GPa, 531K), C (20 GPa, 1387 K), and D (10 GPa, 853.5 K). In addition, (c) shows the phase diagram with temperatures ranging from 0 to 1000 K and pressures ranging from 0 to 3 GPa from the DP-PBE model.  
}}
\label{fig:phase}
\end{figure}

\CT{Figs.~\ref{fig:phase}(a), (b), and (c) illustrate the phase diagram of Sn obtained from the experiment~\cite{14JAP-Sn-xu}, the DP-SCAN, and the DP-PBE model.}
For the DP-SCAN model, we computed the phase diagram with the temperature ranges from 0 to 2000 K while the pressure ranges from 0 to 20 GPa.
%
We first adopted the thermodynamics integration method~\cite{08JPCM-Vega} \MC{(more details can be found in the "Free Energy Calculations" Section of SI~\cite{23SI})} to compute the coexistence points on the phase diagram of Sn, as listed as A, B, and C in Fig.~\ref{fig:phase}(b). 
First, the A point indicates the coexistence point of the $\beta$-Sn and liquid Sn at 5 GPa and 741 K. 
Second, the B point represents the coexistence of the $\beta$-Sn and {\it bct}-Sn phases at 20 GPa and 531 K.
Third, the C point depicts the coexistence of the {\it bct}-Sn and liquid Sn at 20 GPa and 1387 K. 
Finally, we performed the Gibbs-Duhem integration along the coexistence line and reached the $\beta$-{\it bct}-liquid triple-point D at 10 GPa and 853.5 K. 
\CT{Similar routines were performed for the DP-PBE model to yield the phase diagram of Sn, which is shown in Fig.~\ref{fig:phase}(c).}

Although the triple point from the DP-SCAN model quantitatively deviates from the experimental triple-point at 3.02 GPa and 562 K ~\cite{14JAP-Sn-xu}, we emphasize that, to the best of our knowledge, \MC{this is the first time that the triple point of the Sn element can be obtained by molecular dynamics with {\it ab initio} accuracy}, and the resulting phase diagram of the $\beta$-Sn, {\it bct}-Sn, and liquid Sn qualitatively agrees well with the experimental data.
\MC{Notably, in a recent work, Rehn {\it et al.}~\cite{21B-Rehn} performed DFT calculations with the GGA XC functional in the form of AM05~\cite{05B-AM05} to yield parameters for constructing a new model for the Sn phase diagram.
However, the resulting model includes empirical data because the model parameters obtained from DFT results were adjusted to the experimental phase boundaries.}
In particular, the accuracy of the phase diagram largely depends on the exchange-correlation functional and here we used the meta-GGA SCAN functional.  
\CT{For example, Fig.~\ref{fig:phase}(c) illustrates the phase diagram of Sn from the DP-PBE model, which substantially deviates from the experimental phase diagram. The DP-PBE model cannot predict the existence of the $\beta$-{\it bct}-liquid triple-point. Worse still, the DP-PBE model predicts a triple-point of $\alpha$-$\beta$-{\it bct} around 0.6 GPa and 200 K. 
Moreover, we observe the $\alpha$ $\rightarrow$ {\it bct} $\rightarrow$ liquid phase transition around 0 GPa, which is inconsistent with the experiment. 
Therefore, we conclude that the PBE functional cannot accurately describe the phase diagram of Sn.}

\begin{figure}[h!t]
  \begin{center}
    \includegraphics[width=8cm]{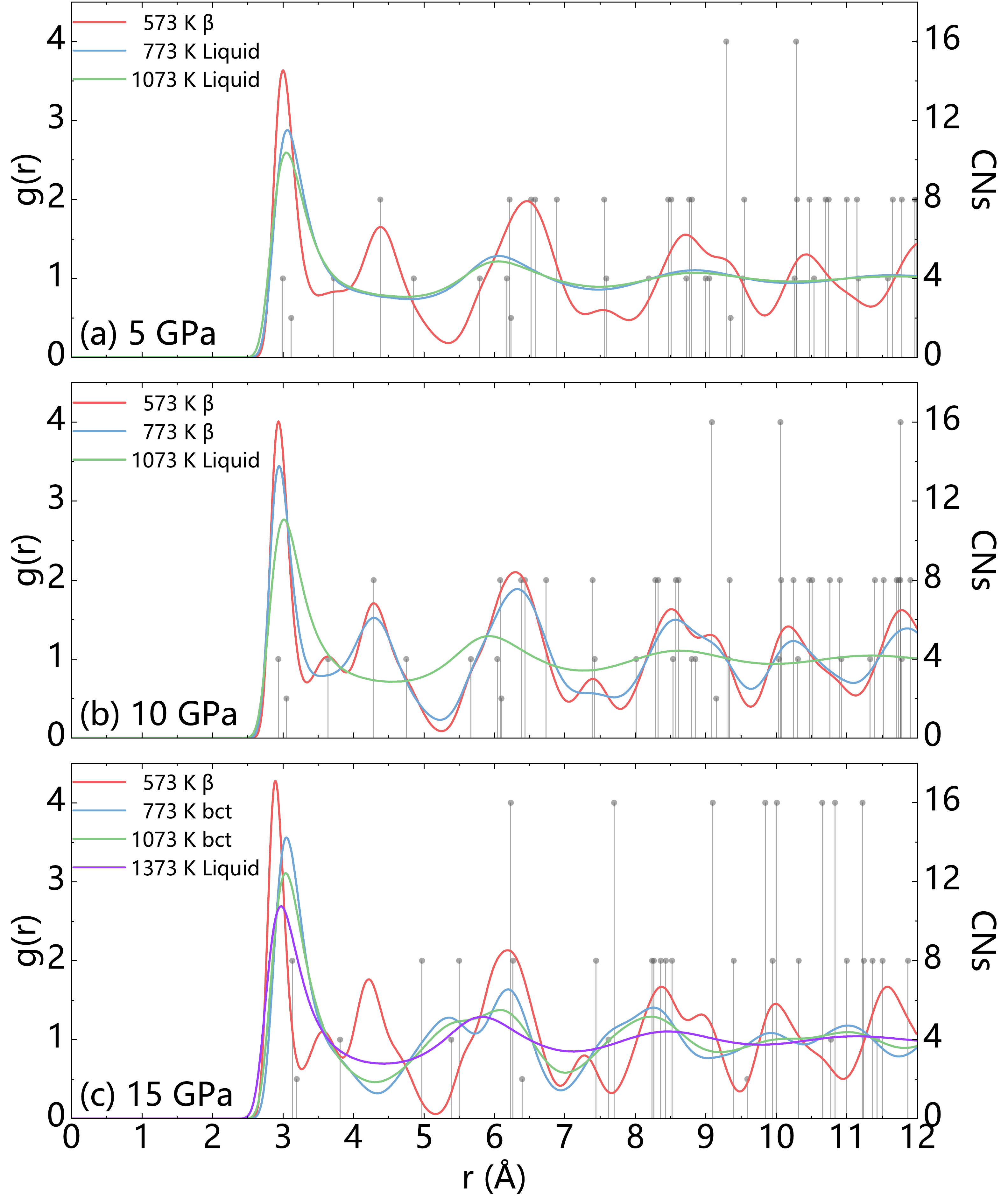}
  \end{center}
\caption{Radial distribution functions $g(r)$ of Sn at different temperature and pressure conditions (listed in TABLE~\ref{table:points}). 
Results of $g(r)$ for the $\beta$-Sn, {\it bct}-Sn, and liquid Sn are shown with the pressure being (a) 5 GPa, (b) 10 GPa, and (c) 15 GPa. 
The gray vertical lines depict the coordination numbers (CNs) of (a) (b) a $\beta$-Sn structure and (c) a {\it bct}-Sn structure. 
The densities of the solid phases are 7.54 $g/cm^3$ in (a), 8.06 $g/cm^3$ in (b) and 8.50 $g/cm^3$ in (c).}
\label{fig:pdf_phase}
\end{figure}

In order to study the detailed structural changes with respect to temperatures and pressures, we chose 10 different temperature and pressure conditions, as listed in Table~\ref{table:points}, to perform MD simulations. 
For each condition, we ran a 250-ps NPT trajectory with a time step of 0.5 fs.
A 2000-atom system was chosen for the liquid-Sn and {\it bct}-Sn systems while a 2048-atom system was set for the $\beta$-Sn phase.
We took $4\times10^4$ snapshots from the last 200-ps trajectory to compute the radial distribution functions $g(r)$.
The results of $g(r)$ for Sn under pressures of 5, 10 and \CT{15} GPa are shown in Figs.~\ref{fig:pdf_phase}(a), (b) and (c), respectively. 
In addition, the coordination numbers (CNs) of a $\beta$-Sn structure and a {\it bct}-Sn structure are shown with vertical gray lines in Figs.~\ref{fig:pdf_phase}(a)(b) and Fig.~\ref{fig:pdf_phase}(c), respectively.

\begin{table}[h!t!b]
\centering
\caption{Conditions (temperature $T$ and pressure $P$) for molecular dynamics simulations of Sn using the DP-SCAN model. 
The $\beta$-Sn, {\it bct}-Sn, and liquid Sn are prepared as initial configurations. 
Three pressures (5, \CT{10, 15} GPa) and four temperatures (573, 773, 1073, and 1373 K) are selected.}
\begin{tabular}{c|cccc}
\hline
        \diagbox {$P$ (GPa)}{$T$ (K)}  & 573 & 773 & 1073 & 1373\\ \hline
        5  & $\beta$ & Liquid     & Liquid    & -      \\
        10 & $\beta$ & $\beta$    & Liquid    & -      \\
        15 & $\beta$ & {\it bct}  & {\it bct} & Liquid \\
\hline
\end{tabular}
\label{table:points}
\end{table}

At a pressure of 5 GPa, the peaks of $g(r)$ at 573 K shown in Fig.~\ref{fig:pdf_phase}(a) are mainly consistent with the peaks from a $\beta$-Sn structure. 
At a higher temperature of 773 K, dramatic changes of $g(r)$ occur, suggesting that a solid-liquid phase transition occurs. 
Specifically, the location of the first peak shifts outwards (from 3.00 to 3.06 {\AA}) while the second peak (4.38 {\AA}) vanishes.
Furthermore, the third peak (6.45 {\AA}) lowers and shifts inwards.
The first peak of $g(r)$ suggests the strong interactions between Sn and its first-shell neighbors, and the peak decreases when the temperature increases to 1073 K, indicating that the high temperature could weaken the strength of the bond.
The disappearance of the second peak and the changes of other peaks suggest that the original structural order in the $\beta$-Sn is destroyed. 
Instead, a new peak at 6.06 {\AA} emerges, indicating that a new local structure forms in the liquid phase. 

When the pressure increases to 10 GPa and the temperature is 573 K, we predict two distinct structural features of the $\beta$-Sn structure located at 3.64 and 7.39 {\AA} of $g(r)$, which are shown in Fig.~\ref{fig:pdf_phase}(b). 
Moreover, a shoulder appears at 9.08 {\AA} near the peak of 8.53 {\AA}. 
Interestingly, the three feature peaks disappear when the temperature increases to 773 K, indicating that the structure of $\beta$-Sn at 10 GPa changes between 573 and 773 K.

Fig.~\ref{fig:pdf_phase}(c) shows that the $\beta$-Sn, {\it bct}-Sn, and liquid Sn structures exist at different temperatures when the pressure further increases to \CT{15} GPa.
At 573 K, the $g(r)$ of $\beta$-Sn is close to the one at 10 GPa. 
However, when the temperature increases to 773 K, the $\beta$-Sn transforms to the {\it bct}-Sn structure. 
As the temperature rises from 773 to 1373 K, the location of the first peak gradually shifts from 3.06 to 2.97 {\AA}, while the second and third peaks of {\it bct}-Sn first merge to a new broad peak and then becomes a narrower peak in liquid Sn. 
Besides, we calculated the coordination numbers by counting the first peak of $g(r)$ up to the first minimum; the resulting data are 6.5 at 573 K, 10.9 at 773 K, 9.4 at 1073 K, and 10.7 at 1373 K. 
This indicates that the local structure of liquid Sn shares some similarities with a distorted {\it bct}-like structure.


\section{Conclusions}
\label{Conclusions}

In this work, we adopted the DP model, i.e., a machine-learning-based many-body potential energy model, to study the properties of solid and liquid Sn at various temperatures and pressures.
A \MC{concurrent} learning scheme was used with the DP-GEN workflow, which automatically generated the training data for Sn with pressures ranging from 0 to 50 GPa and temperatures ranging from 0 to 2000 K.
Notably, the training data were obtained from DFT calculations with the meta-GGA SCAN XC functional, which yields more accurate results for Sn than the PBE XC functional.
Based on the generated DP-SCAN model, the cohesive energies, the cell parameters, the bulk modulus, and the elastic constants of various Sn phases including the $\alpha$, $\beta$, {\it bct} and {\it bcc} phases were systematically investigated. 
\MC{We also computed the phonon dispersions for the $\alpha$ and $\beta$ phases.}

We found the DP-SCAN model not only yielded results close to the experimental findings for $\alpha$-Sn and $\beta$-Sn but also predicted better properties of {\it bct}-Sn such as the cell parameters $c/a$ and the elastic constant $C_{44}$ compared with previous classical potentials.
Meanwhile, the radial distribution functions, ionic densities, and diffusion coefficients for liquid Sn were consistent with the experiment. 
\MC{We also predicted properties of liquid Sn at high pressures.}
In addition, we used the DP-SCAN model to compute the phase boundaries of $\alpha$, $\beta$, {\it bct}, and liquid phases of Sn by comparing the free energies of different phases as obtained from the thermodynamics integration method.
The DP-SCAN model successfully predicted the existence of a triple-point occurring \CT{around} 10 GPa and 853.5 K and yielded the phase diagram of Sn, which qualitatively agreed with the experiment.
We further chose 10 different temperature and pressure conditions to perform MD simulations using the DP-SCAN model and analyzed changes of local structures of Sn across different phases.
%

Due to the sensitivity of the Sn element to external conditions, atomistic simulations of Sn beyond the ambient pressure have been a challenging task for decades.
The current work extended the capability of atomistic simulations of Sn to a wide range of pressures (0-50 GPa) and temperatures (0-2000 K).
Although extending the DP-SCAN potential to higher pressures and temperatures is feasible, it requires more computational resources. 
In this regard, new DP potentials for Sn can be expected in future works.
Importantly, our work demonstrated that the machine-learning-based DP-SCAN model could accurately describe the $\alpha$, $\beta$, {\it bct}, and liquid phases of Sn due to the use of the more accurate SCAN XC functional rather than PBE within the framework of DFT, which opens a new era to provide an atomic-level understanding of the phase diagram of Sn with {\it ab initio} accuracy.
\CT{Notably, the $bct/bco$ mixed phase and the $bco/bcc$ mixed phase of Sn had been found in some recent experiments~\cite{13PRB-Sn-138GPa, 16CPB-Sn, 15PRL-Sn-1.5TPa}.
However, these phases share close energy in a small range of temperatures and pressures, which poses huge challenges for simulations to accurately determine the phase boundaries between the $bct$, $bco$, and $bcc$ phases~\cite{22JAC-Guillaume}. 
Since our focus of the current work is on the $\beta$-$bct$-liquid triple-point, we did not investigate the $bco$ phase in this work. 
It would be worth exploring the phase diagram of $bco$ and $bcc$ in the future.}
In conclusion, our strategy to generate DP models of Sn with the {\it ab initio} accuracy enables the community to study more aspects of Sn at an affordable cost. For instance, the underlying mechanism of phase transitions at extreme conditions and the phase boundaries under shock compression can be explored in future works.

$\\$
{\bf Acknowledgements}
$\\$
The work of M.C. is supported by the National Science Foundation of P. R. China under Grants Nos.~12122401, 12074007, and 12135002.
The work of H.W. is supported by the National Science Foundation of P. R. China under Grants Nos.~11871110 and 12122103.
We thank the Bohrium platform for providing high performance computers for our simulations.
Part of the numerical simulations were performed on the High Performance Computing Platform of CAPT.

$\\$
{\bf AUTHOR DECLARATIONS}
$\\$
{\bf Competing interests}
$\\$
The authors declare no competing interests.


$\\$
{\bf Availability of Data}
$\\$
The DP-SCAN and DP-PBE potentials are attached in the SI, which can be used with the DeePMD-kit package.
The data supporting this study's findings are available from the corresponding author upon reasonable request.

\bibliography{ref}

\begin{thebibliography}{84}
\expandafter\ifx\csname natexlab\endcsname\relax\def\natexlab#1{#1}\fi
\expandafter\ifx\csname bibnamefont\endcsname\relax
  \def\bibnamefont#1{#1}\fi
\expandafter\ifx\csname bibfnamefont\endcsname\relax
  \def\bibfnamefont#1{#1}\fi
\expandafter\ifx\csname citenamefont\endcsname\relax
  \def\citenamefont#1{#1}\fi
\expandafter\ifx\csname url\endcsname\relax
  \def\url#1{\texttt{#1}}\fi
\expandafter\ifx\csname urlprefix\endcsname\relax\def\urlprefix{URL }\fi
\providecommand{\bibinfo}[2]{#2}
\providecommand{\eprint}[2][]{\url{#2}}

\bibitem[{\citenamefont{Zeng and Tu}(2002)}]{Sn-alloy-solder-1}
\bibinfo{author}{\bibfnamefont{K.}~\bibnamefont{Zeng}} \bibnamefont{and}
  \bibinfo{author}{\bibfnamefont{K.}~\bibnamefont{Tu}},
  \bibinfo{journal}{Mater. Sci. Eng. R Rep.} \textbf{\bibinfo{volume}{38}},
  \bibinfo{pages}{55} (\bibinfo{year}{2002}).

\bibitem[{\citenamefont{Miller et~al.}(1994)\citenamefont{Miller, Anderson, and
  Smith}}]{Sn-alloy-solder-2}
\bibinfo{author}{\bibfnamefont{C.~M.} \bibnamefont{Miller}},
  \bibinfo{author}{\bibfnamefont{I.~E.} \bibnamefont{Anderson}},
  \bibnamefont{and} \bibinfo{author}{\bibfnamefont{J.~F.} \bibnamefont{Smith}},
  \bibinfo{journal}{J. Electron. Mater.} \textbf{\bibinfo{volume}{23}},
  \bibinfo{pages}{595} (\bibinfo{year}{1994}).

\bibitem[{\citenamefont{Coenen et~al.}(2014)\citenamefont{Coenen, Temmerman,
  Federici, Philipps, Sergienko, Strohmayer, Terra, Unterberg, Wegener, and den
  Bekerom}}]{14PS-Coenen}
\bibinfo{author}{\bibfnamefont{J.~W.} \bibnamefont{Coenen}},
  \bibinfo{author}{\bibfnamefont{G.~D.} \bibnamefont{Temmerman}},
  \bibinfo{author}{\bibfnamefont{G.}~\bibnamefont{Federici}},
  \bibinfo{author}{\bibfnamefont{V.}~\bibnamefont{Philipps}},
  \bibinfo{author}{\bibfnamefont{G.}~\bibnamefont{Sergienko}},
  \bibinfo{author}{\bibfnamefont{G.}~\bibnamefont{Strohmayer}},
  \bibinfo{author}{\bibfnamefont{A.}~\bibnamefont{Terra}},
  \bibinfo{author}{\bibfnamefont{B.}~\bibnamefont{Unterberg}},
  \bibinfo{author}{\bibfnamefont{T.}~\bibnamefont{Wegener}}, \bibnamefont{and}
  \bibinfo{author}{\bibfnamefont{D.~C. M.~V.} \bibnamefont{den Bekerom}},
  \bibinfo{journal}{Phys. Scr.} \textbf{\bibinfo{volume}{T159}},
  \bibinfo{pages}{014037} (\bibinfo{year}{2014}).

\bibitem[{\citenamefont{Liu et~al.}(2017)\citenamefont{Liu, Zheng, Ren, He, and
  Chen}}]{17JCP-Xiaohui}
\bibinfo{author}{\bibfnamefont{X.}~\bibnamefont{Liu}},
  \bibinfo{author}{\bibfnamefont{D.}~\bibnamefont{Zheng}},
  \bibinfo{author}{\bibfnamefont{X.}~\bibnamefont{Ren}},
  \bibinfo{author}{\bibfnamefont{L.}~\bibnamefont{He}}, \bibnamefont{and}
  \bibinfo{author}{\bibfnamefont{M.}~\bibnamefont{Chen}}, \bibinfo{journal}{J.
  Chem. Phys.} \textbf{\bibinfo{volume}{147}}, \bibinfo{pages}{064505}
  (\bibinfo{year}{2017}).

\bibitem[{\citenamefont{del Rio et~al.}(2018)\citenamefont{del Rio, Chen,
  González, and Carter}}]{18JCP-Beatriz}
\bibinfo{author}{\bibfnamefont{B.~G.} \bibnamefont{del Rio}},
  \bibinfo{author}{\bibfnamefont{M.}~\bibnamefont{Chen}},
  \bibinfo{author}{\bibfnamefont{L.~E.} \bibnamefont{González}},
  \bibnamefont{and} \bibinfo{author}{\bibfnamefont{E.~A.}
  \bibnamefont{Carter}}, \bibinfo{journal}{J. Chem. Phys.}
  \textbf{\bibinfo{volume}{149}}, \bibinfo{pages}{094504}
  (\bibinfo{year}{2018}).

\bibitem[{\citenamefont{Zheng et~al.}(2021)\citenamefont{Zheng, Shen, Chen,
  Ren, and He}}]{21JNM-Daye}
\bibinfo{author}{\bibfnamefont{D.}~\bibnamefont{Zheng}},
  \bibinfo{author}{\bibfnamefont{Z.-X.} \bibnamefont{Shen}},
  \bibinfo{author}{\bibfnamefont{M.}~\bibnamefont{Chen}},
  \bibinfo{author}{\bibfnamefont{X.}~\bibnamefont{Ren}}, \bibnamefont{and}
  \bibinfo{author}{\bibfnamefont{L.}~\bibnamefont{He}}, \bibinfo{journal}{J.
  Nucl. Mater.} \textbf{\bibinfo{volume}{543}}, \bibinfo{pages}{152542}
  (\bibinfo{year}{2021}).

\bibitem[{\citenamefont{Christensen et~al.}(1986)\citenamefont{Christensen,
  Satpathy, and Pawlowska}}]{86B-Pb}
\bibinfo{author}{\bibfnamefont{N.~E.} \bibnamefont{Christensen}},
  \bibinfo{author}{\bibfnamefont{S.}~\bibnamefont{Satpathy}}, \bibnamefont{and}
  \bibinfo{author}{\bibfnamefont{Z.}~\bibnamefont{Pawlowska}},
  \bibinfo{journal}{Phys. Rev. B} \textbf{\bibinfo{volume}{34}},
  \bibinfo{pages}{5977} (\bibinfo{year}{1986}).

\bibitem[{\citenamefont{Katzke et~al.}(2006)\citenamefont{Katzke, Bismayer, and
  Tol\'edano}}]{06B-Pierre}
\bibinfo{author}{\bibfnamefont{H.}~\bibnamefont{Katzke}},
  \bibinfo{author}{\bibfnamefont{U.}~\bibnamefont{Bismayer}}, \bibnamefont{and}
  \bibinfo{author}{\bibfnamefont{P.}~\bibnamefont{Tol\'edano}},
  \bibinfo{journal}{Phys. Rev. B} \textbf{\bibinfo{volume}{73}},
  \bibinfo{pages}{134105} (\bibinfo{year}{2006}).

\bibitem[{\citenamefont{Busch and Kebn}(1960)}]{60SSP-Busch}
\bibinfo{author}{\bibfnamefont{G.}~\bibnamefont{Busch}} \bibnamefont{and}
  \bibinfo{author}{\bibfnamefont{R.}~\bibnamefont{Kebn}},
  \bibinfo{journal}{Solid State Phys.} \textbf{\bibinfo{volume}{11}},
  \bibinfo{pages}{1} (\bibinfo{year}{1960}).

\bibitem[{\citenamefont{Barnett et~al.}(1966)\citenamefont{Barnett, Bean, and
  Hall}}]{66JAP-Sn-100kbar}
\bibinfo{author}{\bibfnamefont{J.~D.} \bibnamefont{Barnett}},
  \bibinfo{author}{\bibfnamefont{V.~E.} \bibnamefont{Bean}}, \bibnamefont{and}
  \bibinfo{author}{\bibfnamefont{H.~T.} \bibnamefont{Hall}},
  \bibinfo{journal}{J. Appl. Phys.} \textbf{\bibinfo{volume}{37}},
  \bibinfo{pages}{875} (\bibinfo{year}{1966}).

\bibitem[{\citenamefont{Weir et~al.}(2012)\citenamefont{Weir, Lipp, Falabella,
  Samudrala, and Vohra}}]{12JAP-Sn-45GPa}
\bibinfo{author}{\bibfnamefont{S.~T.} \bibnamefont{Weir}},
  \bibinfo{author}{\bibfnamefont{M.~J.} \bibnamefont{Lipp}},
  \bibinfo{author}{\bibfnamefont{S.}~\bibnamefont{Falabella}},
  \bibinfo{author}{\bibfnamefont{G.}~\bibnamefont{Samudrala}},
  \bibnamefont{and} \bibinfo{author}{\bibfnamefont{Y.~K.} \bibnamefont{Vohra}},
  \bibinfo{journal}{J. Appl. Phys.} \textbf{\bibinfo{volume}{111}},
  \bibinfo{pages}{123529} (\bibinfo{year}{2012}).

\bibitem[{\citenamefont{Xu et~al.}(2014)\citenamefont{Xu, Bi, Li, Wang, Cao,
  Cai, Wang, and Meng}}]{14JAP-Sn-xu}
\bibinfo{author}{\bibfnamefont{L.}~\bibnamefont{Xu}},
  \bibinfo{author}{\bibfnamefont{Y.}~\bibnamefont{Bi}},
  \bibinfo{author}{\bibfnamefont{X.}~\bibnamefont{Li}},
  \bibinfo{author}{\bibfnamefont{Y.}~\bibnamefont{Wang}},
  \bibinfo{author}{\bibfnamefont{X.}~\bibnamefont{Cao}},
  \bibinfo{author}{\bibfnamefont{L.}~\bibnamefont{Cai}},
  \bibinfo{author}{\bibfnamefont{Z.}~\bibnamefont{Wang}}, \bibnamefont{and}
  \bibinfo{author}{\bibfnamefont{C.}~\bibnamefont{Meng}}, \bibinfo{journal}{J.
  Appl. Phys.} \textbf{\bibinfo{volume}{115}}, \bibinfo{pages}{164903}
  (\bibinfo{year}{2014}).

\bibitem[{\citenamefont{Salamat et~al.}(2011)\citenamefont{Salamat, Garbarino,
  Dewaele, Bouvier, Petitgirard, Pickard, McMillan, and
  Mezouar}}]{11PRB-Sn-157GPa}
\bibinfo{author}{\bibfnamefont{A.}~\bibnamefont{Salamat}},
  \bibinfo{author}{\bibfnamefont{G.}~\bibnamefont{Garbarino}},
  \bibinfo{author}{\bibfnamefont{A.}~\bibnamefont{Dewaele}},
  \bibinfo{author}{\bibfnamefont{P.}~\bibnamefont{Bouvier}},
  \bibinfo{author}{\bibfnamefont{S.}~\bibnamefont{Petitgirard}},
  \bibinfo{author}{\bibfnamefont{C.~J.} \bibnamefont{Pickard}},
  \bibinfo{author}{\bibfnamefont{P.~F.} \bibnamefont{McMillan}},
  \bibnamefont{and} \bibinfo{author}{\bibfnamefont{M.}~\bibnamefont{Mezouar}},
  \bibinfo{journal}{Phys. Rev. B} \textbf{\bibinfo{volume}{84}},
  \bibinfo{pages}{140104} (\bibinfo{year}{2011}).

\bibitem[{\citenamefont{Salamat et~al.}(2013)\citenamefont{Salamat, Briggs,
  Bouvier, Petitgirard, Dewaele, Cutler, Cor\`a, Daisenberger, Garbarino, and
  McMillan}}]{13PRB-Sn-138GPa}
\bibinfo{author}{\bibfnamefont{A.}~\bibnamefont{Salamat}},
  \bibinfo{author}{\bibfnamefont{R.}~\bibnamefont{Briggs}},
  \bibinfo{author}{\bibfnamefont{P.}~\bibnamefont{Bouvier}},
  \bibinfo{author}{\bibfnamefont{S.}~\bibnamefont{Petitgirard}},
  \bibinfo{author}{\bibfnamefont{A.}~\bibnamefont{Dewaele}},
  \bibinfo{author}{\bibfnamefont{M.~E.} \bibnamefont{Cutler}},
  \bibinfo{author}{\bibfnamefont{F.}~\bibnamefont{Cor\`a}},
  \bibinfo{author}{\bibfnamefont{D.}~\bibnamefont{Daisenberger}},
  \bibinfo{author}{\bibfnamefont{G.}~\bibnamefont{Garbarino}},
  \bibnamefont{and} \bibinfo{author}{\bibfnamefont{P.~F.}
  \bibnamefont{McMillan}}, \bibinfo{journal}{Phys. Rev. B}
  \textbf{\bibinfo{volume}{88}}, \bibinfo{pages}{104104}
  (\bibinfo{year}{2013}).

\bibitem[{\citenamefont{Jing et~al.}(2016)\citenamefont{Jing, Cao, Zhang, Li,
  He, Hou, Liu, Liu, Bi, Geng et~al.}}]{16CPB-Sn}
\bibinfo{author}{\bibfnamefont{Q.-M.} \bibnamefont{Jing}},
  \bibinfo{author}{\bibfnamefont{Y.-H.} \bibnamefont{Cao}},
  \bibinfo{author}{\bibfnamefont{Y.}~\bibnamefont{Zhang}},
  \bibinfo{author}{\bibfnamefont{S.-R.} \bibnamefont{Li}},
  \bibinfo{author}{\bibfnamefont{Q.}~\bibnamefont{He}},
  \bibinfo{author}{\bibfnamefont{Q.-Y.} \bibnamefont{Hou}},
  \bibinfo{author}{\bibfnamefont{S.-G.} \bibnamefont{Liu}},
  \bibinfo{author}{\bibfnamefont{L.}~\bibnamefont{Liu}},
  \bibinfo{author}{\bibfnamefont{Y.}~\bibnamefont{Bi}},
  \bibinfo{author}{\bibfnamefont{H.-Y.} \bibnamefont{Geng}},
  \bibnamefont{et~al.}, \bibinfo{journal}{Chin. Phys. B}
  \textbf{\bibinfo{volume}{25}}, \bibinfo{pages}{120702}
  (\bibinfo{year}{2016}).

\bibitem[{\citenamefont{Lazicki et~al.}(2015)\citenamefont{Lazicki, Rygg,
  Coppari, Smith, Fratanduono, Kraus, Collins, Briggs, Braun, Swift
  et~al.}}]{15PRL-Sn-1.5TPa}
\bibinfo{author}{\bibfnamefont{A.}~\bibnamefont{Lazicki}},
  \bibinfo{author}{\bibfnamefont{J.~R.} \bibnamefont{Rygg}},
  \bibinfo{author}{\bibfnamefont{F.}~\bibnamefont{Coppari}},
  \bibinfo{author}{\bibfnamefont{R.}~\bibnamefont{Smith}},
  \bibinfo{author}{\bibfnamefont{D.}~\bibnamefont{Fratanduono}},
  \bibinfo{author}{\bibfnamefont{R.~G.} \bibnamefont{Kraus}},
  \bibinfo{author}{\bibfnamefont{G.~W.} \bibnamefont{Collins}},
  \bibinfo{author}{\bibfnamefont{R.}~\bibnamefont{Briggs}},
  \bibinfo{author}{\bibfnamefont{D.~G.} \bibnamefont{Braun}},
  \bibinfo{author}{\bibfnamefont{D.~C.} \bibnamefont{Swift}},
  \bibnamefont{et~al.}, \bibinfo{journal}{Phys. Rev. Lett.}
  \textbf{\bibinfo{volume}{115}}, \bibinfo{pages}{075502}
  (\bibinfo{year}{2015}).

\bibitem[{\citenamefont{La~Lone et~al.}(2019)\citenamefont{La~Lone, Asimow,
  Fat’yanov, Hixson, Stevens, Turley, and Veeser}}]{19JAP-Lone}
\bibinfo{author}{\bibfnamefont{B.~M.} \bibnamefont{La~Lone}},
  \bibinfo{author}{\bibfnamefont{P.~D.} \bibnamefont{Asimow}},
  \bibinfo{author}{\bibfnamefont{O.~V.} \bibnamefont{Fat’yanov}},
  \bibinfo{author}{\bibfnamefont{R.~S.} \bibnamefont{Hixson}},
  \bibinfo{author}{\bibfnamefont{G.~D.} \bibnamefont{Stevens}},
  \bibinfo{author}{\bibfnamefont{W.~D.} \bibnamefont{Turley}},
  \bibnamefont{and} \bibinfo{author}{\bibfnamefont{L.~R.}
  \bibnamefont{Veeser}}, \bibinfo{journal}{J. Appl. Phys.}
  \textbf{\bibinfo{volume}{126}}, \bibinfo{pages}{225103}
  (\bibinfo{year}{2019}).

\bibitem[{\citenamefont{Briggs et~al.}(2019)\citenamefont{Briggs, Torchio,
  Sollier, Occelli, Videau, Kretzschmar, and Wulff}}]{19JSR-Briggs}
\bibinfo{author}{\bibfnamefont{R.}~\bibnamefont{Briggs}},
  \bibinfo{author}{\bibfnamefont{R.}~\bibnamefont{Torchio}},
  \bibinfo{author}{\bibfnamefont{A.}~\bibnamefont{Sollier}},
  \bibinfo{author}{\bibfnamefont{F.}~\bibnamefont{Occelli}},
  \bibinfo{author}{\bibfnamefont{L.}~\bibnamefont{Videau}},
  \bibinfo{author}{\bibfnamefont{N.}~\bibnamefont{Kretzschmar}},
  \bibnamefont{and} \bibinfo{author}{\bibfnamefont{M.}~\bibnamefont{Wulff}},
  \bibinfo{journal}{J. Synchrotron Radiat.} \textbf{\bibinfo{volume}{26}},
  \bibinfo{pages}{96} (\bibinfo{year}{2019}).

\bibitem[{\citenamefont{Deffrennes et~al.}(2022)\citenamefont{Deffrennes,
  Faure, Bottin, Joubert, and Oudot}}]{22JAC-Guillaume}
\bibinfo{author}{\bibfnamefont{G.}~\bibnamefont{Deffrennes}},
  \bibinfo{author}{\bibfnamefont{P.}~\bibnamefont{Faure}},
  \bibinfo{author}{\bibfnamefont{F.}~\bibnamefont{Bottin}},
  \bibinfo{author}{\bibfnamefont{J.-M.} \bibnamefont{Joubert}},
  \bibnamefont{and} \bibinfo{author}{\bibfnamefont{B.}~\bibnamefont{Oudot}},
  \bibinfo{journal}{J. Alloys Compd.} \textbf{\bibinfo{volume}{919}},
  \bibinfo{pages}{165675} (\bibinfo{year}{2022}).

\bibitem[{\citenamefont{Hohenberg and Kohn}(1964)}]{64PR-Hohenberg}
\bibinfo{author}{\bibfnamefont{P.}~\bibnamefont{Hohenberg}} \bibnamefont{and}
  \bibinfo{author}{\bibfnamefont{W.}~\bibnamefont{Kohn}},
  \bibinfo{journal}{Phys. Rev.} \textbf{\bibinfo{volume}{136}},
  \bibinfo{pages}{B864} (\bibinfo{year}{1964}).

\bibitem[{\citenamefont{Kohn and Sham}(1965)}]{65PR-Kohn}
\bibinfo{author}{\bibfnamefont{W.}~\bibnamefont{Kohn}} \bibnamefont{and}
  \bibinfo{author}{\bibfnamefont{L.~J.} \bibnamefont{Sham}},
  \bibinfo{journal}{Phys. Rev.} \textbf{\bibinfo{volume}{140}},
  \bibinfo{pages}{A1133} (\bibinfo{year}{1965}).

\bibitem[{\citenamefont{Cheong and Chang}(1991)}]{91PRB-Sn}
\bibinfo{author}{\bibfnamefont{B.~H.} \bibnamefont{Cheong}} \bibnamefont{and}
  \bibinfo{author}{\bibfnamefont{K.~J.} \bibnamefont{Chang}},
  \bibinfo{journal}{Phys. Rev. B} \textbf{\bibinfo{volume}{44}},
  \bibinfo{pages}{4103} (\bibinfo{year}{1991}).

\bibitem[{\citenamefont{Christensen and Methfessel}(1993)}]{93PRB-LDA-Sn}
\bibinfo{author}{\bibfnamefont{N.~E.} \bibnamefont{Christensen}}
  \bibnamefont{and}
  \bibinfo{author}{\bibfnamefont{M.}~\bibnamefont{Methfessel}},
  \bibinfo{journal}{Phys. Rev. B} \textbf{\bibinfo{volume}{48}},
  \bibinfo{pages}{5797} (\bibinfo{year}{1993}).

\bibitem[{\citenamefont{Aguado}(2003)}]{03PRB-LDA-Sn}
\bibinfo{author}{\bibfnamefont{A.}~\bibnamefont{Aguado}},
  \bibinfo{journal}{Phys. Rev. B} \textbf{\bibinfo{volume}{67}},
  \bibinfo{pages}{212104} (\bibinfo{year}{2003}).

\bibitem[{\citenamefont{Cui et~al.}(2008)\citenamefont{Cui, Cai, Feng, Hu,
  Wang, and Wang}}]{08SSC-GGA-Sn}
\bibinfo{author}{\bibfnamefont{S.}~\bibnamefont{Cui}},
  \bibinfo{author}{\bibfnamefont{L.}~\bibnamefont{Cai}},
  \bibinfo{author}{\bibfnamefont{W.}~\bibnamefont{Feng}},
  \bibinfo{author}{\bibfnamefont{H.}~\bibnamefont{Hu}},
  \bibinfo{author}{\bibfnamefont{C.}~\bibnamefont{Wang}}, \bibnamefont{and}
  \bibinfo{author}{\bibfnamefont{Y.}~\bibnamefont{Wang}},
  \bibinfo{journal}{Phys. Status Solidi B Basic Res.}
  \textbf{\bibinfo{volume}{245}}, \bibinfo{pages}{53} (\bibinfo{year}{2008}).

\bibitem[{\citenamefont{Shahi et~al.}(2018)\citenamefont{Shahi, Sun, and
  Perdew}}]{18B-Perdew}
\bibinfo{author}{\bibfnamefont{C.}~\bibnamefont{Shahi}},
  \bibinfo{author}{\bibfnamefont{J.}~\bibnamefont{Sun}}, \bibnamefont{and}
  \bibinfo{author}{\bibfnamefont{J.~P.} \bibnamefont{Perdew}},
  \bibinfo{journal}{Phys. Rev. B} \textbf{\bibinfo{volume}{97}},
  \bibinfo{pages}{094111} (\bibinfo{year}{2018}).

\bibitem[{\citenamefont{Corkill et~al.}(1991)\citenamefont{Corkill, Garca, and
  Cohen}}]{91B-Sn}
\bibinfo{author}{\bibfnamefont{J.~L.} \bibnamefont{Corkill}},
  \bibinfo{author}{\bibfnamefont{A.}~\bibnamefont{Garca}}, \bibnamefont{and}
  \bibinfo{author}{\bibfnamefont{M.~L.} \bibnamefont{Cohen}},
  \bibinfo{journal}{Phys. Rev. B} \textbf{\bibinfo{volume}{43}},
  \bibinfo{pages}{9251} (\bibinfo{year}{1991}).

\bibitem[{\citenamefont{Soulard et~al.}(2022)\citenamefont{Soulard, Carrard,
  and Durand}}]{22JAP-Soulard}
\bibinfo{author}{\bibfnamefont{L.}~\bibnamefont{Soulard}},
  \bibinfo{author}{\bibfnamefont{T.}~\bibnamefont{Carrard}}, \bibnamefont{and}
  \bibinfo{author}{\bibfnamefont{O.}~\bibnamefont{Durand}},
  \bibinfo{journal}{J. Appl. Phys.} \textbf{\bibinfo{volume}{131}},
  \bibinfo{pages}{135901} (\bibinfo{year}{2022}).

\bibitem[{\citenamefont{Yang et~al.}(2022)\citenamefont{Yang, Zhao, Gao, Chen,
  Zeng, and Wang}}]{22JAP-Yang}
\bibinfo{author}{\bibfnamefont{X.}~\bibnamefont{Yang}},
  \bibinfo{author}{\bibfnamefont{H.}~\bibnamefont{Zhao}},
  \bibinfo{author}{\bibfnamefont{X.}~\bibnamefont{Gao}},
  \bibinfo{author}{\bibfnamefont{Z.}~\bibnamefont{Chen}},
  \bibinfo{author}{\bibfnamefont{X.}~\bibnamefont{Zeng}}, \bibnamefont{and}
  \bibinfo{author}{\bibfnamefont{F.}~\bibnamefont{Wang}}, \bibinfo{journal}{J.
  Appl. Phys.} \textbf{\bibinfo{volume}{132}}, \bibinfo{pages}{075903}
  (\bibinfo{year}{2022}).

\bibitem[{\citenamefont{Ravelo and Baskes}(1997)}]{97L-Sn-meam}
\bibinfo{author}{\bibfnamefont{R.}~\bibnamefont{Ravelo}} \bibnamefont{and}
  \bibinfo{author}{\bibfnamefont{M.}~\bibnamefont{Baskes}},
  \bibinfo{journal}{Phys. Rev. Lett.} \textbf{\bibinfo{volume}{79}},
  \bibinfo{pages}{2482} (\bibinfo{year}{1997}).

\bibitem[{\citenamefont{Vella et~al.}(2017)\citenamefont{Vella, Chen,
  Stillinger, Carter, Debenedetti, and Panagiotopoulos}}]{17B-VellaChen-meam}
\bibinfo{author}{\bibfnamefont{J.~R.} \bibnamefont{Vella}},
  \bibinfo{author}{\bibfnamefont{M.}~\bibnamefont{Chen}},
  \bibinfo{author}{\bibfnamefont{F.~H.} \bibnamefont{Stillinger}},
  \bibinfo{author}{\bibfnamefont{E.~A.} \bibnamefont{Carter}},
  \bibinfo{author}{\bibfnamefont{P.~G.} \bibnamefont{Debenedetti}},
  \bibnamefont{and} \bibinfo{author}{\bibfnamefont{A.~Z.}
  \bibnamefont{Panagiotopoulos}}, \bibinfo{journal}{Phys. Rev. B}
  \textbf{\bibinfo{volume}{95}}, \bibinfo{pages}{064202}
  (\bibinfo{year}{2017}).

\bibitem[{\citenamefont{Etesami et~al.}(2018)\citenamefont{Etesami, Baskes,
  Laradji, and Asadi}}]{18AM-PbSn-meam}
\bibinfo{author}{\bibfnamefont{S.~A.} \bibnamefont{Etesami}},
  \bibinfo{author}{\bibfnamefont{M.~I.} \bibnamefont{Baskes}},
  \bibinfo{author}{\bibfnamefont{M.}~\bibnamefont{Laradji}}, \bibnamefont{and}
  \bibinfo{author}{\bibfnamefont{E.}~\bibnamefont{Asadi}},
  \bibinfo{journal}{Acta Mater.} \textbf{\bibinfo{volume}{161}},
  \bibinfo{pages}{320 } (\bibinfo{year}{2018}).

\bibitem[{\citenamefont{Ko et~al.}(2018)\citenamefont{Ko, Kim, Kwon, and
  Lee}}]{18Metals-Sn-meam}
\bibinfo{author}{\bibfnamefont{W.-S.} \bibnamefont{Ko}},
  \bibinfo{author}{\bibfnamefont{D.-H.} \bibnamefont{Kim}},
  \bibinfo{author}{\bibfnamefont{Y.-J.} \bibnamefont{Kwon}}, \bibnamefont{and}
  \bibinfo{author}{\bibfnamefont{M.}~\bibnamefont{Lee}},
  \bibinfo{journal}{Metals} \textbf{\bibinfo{volume}{8}}, \bibinfo{pages}{900}
  (\bibinfo{year}{2018}).

\bibitem[{\citenamefont{Lee and Baskes}(2000)}]{00B-2NNmeam}
\bibinfo{author}{\bibfnamefont{B.-J.} \bibnamefont{Lee}} \bibnamefont{and}
  \bibinfo{author}{\bibfnamefont{M.~I.} \bibnamefont{Baskes}},
  \bibinfo{journal}{Phys. Rev. B} \textbf{\bibinfo{volume}{62}},
  \bibinfo{pages}{8564} (\bibinfo{year}{2000}).

\bibitem[{\citenamefont{Lee et~al.}(2001)\citenamefont{Lee, Baskes, Kim, and
  Koo~Cho}}]{01B-2NNmeam-bcc}
\bibinfo{author}{\bibfnamefont{B.-J.} \bibnamefont{Lee}},
  \bibinfo{author}{\bibfnamefont{M.}~\bibnamefont{Baskes}},
  \bibinfo{author}{\bibfnamefont{H.}~\bibnamefont{Kim}}, \bibnamefont{and}
  \bibinfo{author}{\bibfnamefont{Y.}~\bibnamefont{Koo~Cho}},
  \bibinfo{journal}{Phys. Rev. B} \textbf{\bibinfo{volume}{64}},
  \bibinfo{pages}{184102} (\bibinfo{year}{2001}).

\bibitem[{\citenamefont{Lee et~al.}(2010)\citenamefont{Lee, Ko, Kim, and
  Kim}}]{10Calphad-2NNmeam}
\bibinfo{author}{\bibfnamefont{B.-J.} \bibnamefont{Lee}},
  \bibinfo{author}{\bibfnamefont{W.-S.} \bibnamefont{Ko}},
  \bibinfo{author}{\bibfnamefont{H.-K.} \bibnamefont{Kim}}, \bibnamefont{and}
  \bibinfo{author}{\bibfnamefont{E.-H.} \bibnamefont{Kim}},
  \bibinfo{journal}{CALPHAD} \textbf{\bibinfo{volume}{34}}, \bibinfo{pages}{510
  } (\bibinfo{year}{2010}).

\bibitem[{\citenamefont{Behler}(2011)}]{11PCCP-Behler}
\bibinfo{author}{\bibfnamefont{J.}~\bibnamefont{Behler}},
  \bibinfo{journal}{Phys. Chem. Chem. Phys.} \textbf{\bibinfo{volume}{13}},
  \bibinfo{pages}{17930} (\bibinfo{year}{2011}).

\bibitem[{\citenamefont{Bartók and Csányi}(2015)}]{15QC-GAP}
\bibinfo{author}{\bibfnamefont{A.~P.} \bibnamefont{Bartók}} \bibnamefont{and}
  \bibinfo{author}{\bibfnamefont{G.}~\bibnamefont{Csányi}},
  \bibinfo{journal}{Int. J. Quantum Chemistry} \textbf{\bibinfo{volume}{115}},
  \bibinfo{pages}{1051} (\bibinfo{year}{2015}).

\bibitem[{\citenamefont{Han et~al.}(2021)\citenamefont{Han, Chen, Wang, Chen,
  Xu, Wu, Chen, Lu, and Guan}}]{21CMS-Lihong}
\bibinfo{author}{\bibfnamefont{L.}~\bibnamefont{Han}},
  \bibinfo{author}{\bibfnamefont{X.}~\bibnamefont{Chen}},
  \bibinfo{author}{\bibfnamefont{Q.}~\bibnamefont{Wang}},
  \bibinfo{author}{\bibfnamefont{Y.}~\bibnamefont{Chen}},
  \bibinfo{author}{\bibfnamefont{M.}~\bibnamefont{Xu}},
  \bibinfo{author}{\bibfnamefont{L.}~\bibnamefont{Wu}},
  \bibinfo{author}{\bibfnamefont{C.}~\bibnamefont{Chen}},
  \bibinfo{author}{\bibfnamefont{P.}~\bibnamefont{Lu}}, \bibnamefont{and}
  \bibinfo{author}{\bibfnamefont{P.}~\bibnamefont{Guan}},
  \bibinfo{journal}{Comput. Mater. Sci.} \textbf{\bibinfo{volume}{200}},
  \bibinfo{pages}{110829} (\bibinfo{year}{2021}).

\bibitem[{\citenamefont{Zhang et~al.}(2018)\citenamefont{Zhang, Han, Wang, Car,
  and E}}]{18PRL-deepmd}
\bibinfo{author}{\bibfnamefont{L.}~\bibnamefont{Zhang}},
  \bibinfo{author}{\bibfnamefont{J.}~\bibnamefont{Han}},
  \bibinfo{author}{\bibfnamefont{H.}~\bibnamefont{Wang}},
  \bibinfo{author}{\bibfnamefont{R.}~\bibnamefont{Car}}, \bibnamefont{and}
  \bibinfo{author}{\bibfnamefont{W.}~\bibnamefont{E}}, \bibinfo{journal}{Phys.
  Rev. Lett.} \textbf{\bibinfo{volume}{120}}, \bibinfo{pages}{143001}
  (\bibinfo{year}{2018}).

\bibitem[{\citenamefont{Wang et~al.}(2018)\citenamefont{Wang, Zhang, Han, and
  E}}]{18CPC-deepmd}
\bibinfo{author}{\bibfnamefont{H.}~\bibnamefont{Wang}},
  \bibinfo{author}{\bibfnamefont{L.}~\bibnamefont{Zhang}},
  \bibinfo{author}{\bibfnamefont{J.}~\bibnamefont{Han}}, \bibnamefont{and}
  \bibinfo{author}{\bibfnamefont{W.}~\bibnamefont{E}},
  \bibinfo{journal}{Comput. Phys. Commun.} \textbf{\bibinfo{volume}{228}},
  \bibinfo{pages}{178 } (\bibinfo{year}{2018}).

\bibitem[{\citenamefont{Jiang et~al.}(2021)\citenamefont{Jiang, Zhang, Zhang,
  and Wang}}]{21CPB}
\bibinfo{author}{\bibfnamefont{W.}~\bibnamefont{Jiang}},
  \bibinfo{author}{\bibfnamefont{Y.}~\bibnamefont{Zhang}},
  \bibinfo{author}{\bibfnamefont{L.}~\bibnamefont{Zhang}}, \bibnamefont{and}
  \bibinfo{author}{\bibfnamefont{H.}~\bibnamefont{Wang}},
  \bibinfo{journal}{Chin. Phys. B} \textbf{\bibinfo{volume}{30}},
  \bibinfo{pages}{050706} (\bibinfo{year}{2021}).

\bibitem[{\citenamefont{Niu et~al.}(2020)\citenamefont{Niu, Bonati, Piaggi, and
  Parrinello}}]{20NC-Niu}
\bibinfo{author}{\bibfnamefont{H.}~\bibnamefont{Niu}},
  \bibinfo{author}{\bibfnamefont{L.}~\bibnamefont{Bonati}},
  \bibinfo{author}{\bibfnamefont{P.~M.} \bibnamefont{Piaggi}},
  \bibnamefont{and}
  \bibinfo{author}{\bibfnamefont{M.}~\bibnamefont{Parrinello}},
  \bibinfo{journal}{Nat. Commun.} \textbf{\bibinfo{volume}{11}},
  \bibinfo{pages}{2654} (\bibinfo{year}{2020}).

\bibitem[{\citenamefont{Zhang et~al.}(2021)\citenamefont{Zhang, Wang, Car, and
  E}}]{21L-LinFeng}
\bibinfo{author}{\bibfnamefont{L.}~\bibnamefont{Zhang}},
  \bibinfo{author}{\bibfnamefont{H.}~\bibnamefont{Wang}},
  \bibinfo{author}{\bibfnamefont{R.}~\bibnamefont{Car}}, \bibnamefont{and}
  \bibinfo{author}{\bibfnamefont{W.}~\bibnamefont{E}}, \bibinfo{journal}{Phys.
  Rev. Lett.} \textbf{\bibinfo{volume}{126}}, \bibinfo{pages}{236001}
  (\bibinfo{year}{2021}).

\bibitem[{\citenamefont{Xu et~al.}(2020)\citenamefont{Xu, Zhang, Zhang, Chen,
  Santra, and Wu}}]{20PRB-Jianhang}
\bibinfo{author}{\bibfnamefont{J.}~\bibnamefont{Xu}},
  \bibinfo{author}{\bibfnamefont{C.}~\bibnamefont{Zhang}},
  \bibinfo{author}{\bibfnamefont{L.}~\bibnamefont{Zhang}},
  \bibinfo{author}{\bibfnamefont{M.}~\bibnamefont{Chen}},
  \bibinfo{author}{\bibfnamefont{B.}~\bibnamefont{Santra}}, \bibnamefont{and}
  \bibinfo{author}{\bibfnamefont{X.}~\bibnamefont{Wu}}, \bibinfo{journal}{Phys.
  Rev. B} \textbf{\bibinfo{volume}{102}}, \bibinfo{pages}{214113}
  (\bibinfo{year}{2020}).

\bibitem[{\citenamefont{Liu et~al.}(2020)\citenamefont{Liu, Lu, and
  Chen}}]{20JPCM-Qianrui}
\bibinfo{author}{\bibfnamefont{Q.}~\bibnamefont{Liu}},
  \bibinfo{author}{\bibfnamefont{D.}~\bibnamefont{Lu}}, \bibnamefont{and}
  \bibinfo{author}{\bibfnamefont{M.}~\bibnamefont{Chen}}, \bibinfo{journal}{J.
  Phys. Condens. Matter} \textbf{\bibinfo{volume}{32}}, \bibinfo{pages}{144002}
  (\bibinfo{year}{2020}).

\bibitem[{\citenamefont{Liu et~al.}(2021)\citenamefont{Liu, Li, and
  Chen}}]{21MRE}
\bibinfo{author}{\bibfnamefont{Q.}~\bibnamefont{Liu}},
  \bibinfo{author}{\bibfnamefont{J.}~\bibnamefont{Li}}, \bibnamefont{and}
  \bibinfo{author}{\bibfnamefont{M.}~\bibnamefont{Chen}},
  \bibinfo{journal}{Matter Radiat. at Extremes} \textbf{\bibinfo{volume}{6}},
  \bibinfo{pages}{026902} (\bibinfo{year}{2021}).

\bibitem[{\citenamefont{Lu et~al.}(2021)\citenamefont{Lu, Wang, Chen, Lin, Car,
  E, Jia, and Zhang}}]{20CPC-Denghui}
\bibinfo{author}{\bibfnamefont{D.}~\bibnamefont{Lu}},
  \bibinfo{author}{\bibfnamefont{H.}~\bibnamefont{Wang}},
  \bibinfo{author}{\bibfnamefont{M.}~\bibnamefont{Chen}},
  \bibinfo{author}{\bibfnamefont{L.}~\bibnamefont{Lin}},
  \bibinfo{author}{\bibfnamefont{R.}~\bibnamefont{Car}},
  \bibinfo{author}{\bibfnamefont{W.}~\bibnamefont{E}},
  \bibinfo{author}{\bibfnamefont{W.}~\bibnamefont{Jia}}, \bibnamefont{and}
  \bibinfo{author}{\bibfnamefont{L.}~\bibnamefont{Zhang}},
  \bibinfo{journal}{Comput. Phys. Commun.} \textbf{\bibinfo{volume}{259}},
  \bibinfo{pages}{107624} (\bibinfo{year}{2021}).

\bibitem[{\citenamefont{Jia et~al.}(2020)\citenamefont{Jia, Wang, Chen, Lu,
  Lin, Car, E, and Zhang}}]{20SC-Weile}
\bibinfo{author}{\bibfnamefont{W.}~\bibnamefont{Jia}},
  \bibinfo{author}{\bibfnamefont{H.}~\bibnamefont{Wang}},
  \bibinfo{author}{\bibfnamefont{M.}~\bibnamefont{Chen}},
  \bibinfo{author}{\bibfnamefont{D.}~\bibnamefont{Lu}},
  \bibinfo{author}{\bibfnamefont{L.}~\bibnamefont{Lin}},
  \bibinfo{author}{\bibfnamefont{R.}~\bibnamefont{Car}},
  \bibinfo{author}{\bibfnamefont{W.}~\bibnamefont{E}}, \bibnamefont{and}
  \bibinfo{author}{\bibfnamefont{L.}~\bibnamefont{Zhang}}, in
  \emph{\bibinfo{booktitle}{Proceedings of the International Conference for
  High Performance Computing, Networking, Storage and Analysis}}
  (\bibinfo{publisher}{IEEE Press}, \bibinfo{year}{2020}), SC '20.

\bibitem[{\citenamefont{Kresse and Furthm\"uller}(1996)}]{96B-VASP}
\bibinfo{author}{\bibfnamefont{G.}~\bibnamefont{Kresse}} \bibnamefont{and}
  \bibinfo{author}{\bibfnamefont{J.}~\bibnamefont{Furthm\"uller}},
  \bibinfo{journal}{Phys. Rev. B} \textbf{\bibinfo{volume}{54}},
  \bibinfo{pages}{11169} (\bibinfo{year}{1996}).

\bibitem[{\citenamefont{Bl\"ochl}(1994)}]{94B-PAW}
\bibinfo{author}{\bibfnamefont{P.~E.} \bibnamefont{Bl\"ochl}},
  \bibinfo{journal}{Phys. Rev. B} \textbf{\bibinfo{volume}{50}},
  \bibinfo{pages}{17953} (\bibinfo{year}{1994}).

\bibitem[{\citenamefont{Kresse and Joubert}(1999)}]{99B-USPP}
\bibinfo{author}{\bibfnamefont{G.}~\bibnamefont{Kresse}} \bibnamefont{and}
  \bibinfo{author}{\bibfnamefont{D.}~\bibnamefont{Joubert}},
  \bibinfo{journal}{Phys. Rev. B} \textbf{\bibinfo{volume}{59}},
  \bibinfo{pages}{1758} (\bibinfo{year}{1999}).

\bibitem[{\citenamefont{Sun et~al.}(2015)\citenamefont{Sun, Ruzsinszky, and
  Perdew}}]{15PRL-SCAN}
\bibinfo{author}{\bibfnamefont{J.}~\bibnamefont{Sun}},
  \bibinfo{author}{\bibfnamefont{A.}~\bibnamefont{Ruzsinszky}},
  \bibnamefont{and} \bibinfo{author}{\bibfnamefont{J.~P.}
  \bibnamefont{Perdew}}, \bibinfo{journal}{Phys. Rev. Lett.}
  \textbf{\bibinfo{volume}{115}}, \bibinfo{pages}{036402}
  (\bibinfo{year}{2015}).

\bibitem[{\citenamefont{Monkhorst and Pack}(1976)}]{76B-MP}
\bibinfo{author}{\bibfnamefont{H.~J.} \bibnamefont{Monkhorst}}
  \bibnamefont{and} \bibinfo{author}{\bibfnamefont{J.~D.} \bibnamefont{Pack}},
  \bibinfo{journal}{Phys. Rev. B} \textbf{\bibinfo{volume}{13}},
  \bibinfo{pages}{5188} (\bibinfo{year}{1976}).

\bibitem[{\citenamefont{Kingma and Ba}(2017)}]{17arX-Adam}
\bibinfo{author}{\bibfnamefont{D.~P.} \bibnamefont{Kingma}} \bibnamefont{and}
  \bibinfo{author}{\bibfnamefont{J.}~\bibnamefont{Ba}},
  \bibinfo{journal}{arXiv:1412.6980}  (\bibinfo{year}{2017}).

\bibitem[{\citenamefont{Zhang et~al.}(2019)\citenamefont{Zhang, Lin, Wang, Car,
  and E}}]{19PRM-DPGEN}
\bibinfo{author}{\bibfnamefont{L.}~\bibnamefont{Zhang}},
  \bibinfo{author}{\bibfnamefont{D.-Y.} \bibnamefont{Lin}},
  \bibinfo{author}{\bibfnamefont{H.}~\bibnamefont{Wang}},
  \bibinfo{author}{\bibfnamefont{R.}~\bibnamefont{Car}}, \bibnamefont{and}
  \bibinfo{author}{\bibfnamefont{W.}~\bibnamefont{E}}, \bibinfo{journal}{Phys.
  Rev. Mater.} \textbf{\bibinfo{volume}{3}}, \bibinfo{pages}{023804}
  (\bibinfo{year}{2019}).

\bibitem[{\citenamefont{Zhang et~al.}(2020)\citenamefont{Zhang, Wang, Chen,
  Zeng, Zhang, Wang, and E}}]{20CPC-DPGEN}
\bibinfo{author}{\bibfnamefont{Y.}~\bibnamefont{Zhang}},
  \bibinfo{author}{\bibfnamefont{H.}~\bibnamefont{Wang}},
  \bibinfo{author}{\bibfnamefont{W.}~\bibnamefont{Chen}},
  \bibinfo{author}{\bibfnamefont{J.}~\bibnamefont{Zeng}},
  \bibinfo{author}{\bibfnamefont{L.}~\bibnamefont{Zhang}},
  \bibinfo{author}{\bibfnamefont{H.}~\bibnamefont{Wang}}, \bibnamefont{and}
  \bibinfo{author}{\bibfnamefont{W.}~\bibnamefont{E}},
  \bibinfo{journal}{Comput. Phys. Commun.} \textbf{\bibinfo{volume}{253}},
  \bibinfo{pages}{107206} (\bibinfo{year}{2020}).

\bibitem[{23S()}]{23SI}
\bibinfo{note}{See Supporting Information at [URL] for detailed setups of the
  exploration strategy listed in Table S1; the bulk properties of solid Sn
  phases from the PBE functional listed in Table S2; the coordinates of special
  $k$-points listed in Tables S3 and S4; details of Free Energy Calculations;
  input file of DeePMD-Kit.}

\bibitem[{\citenamefont{Plimpton}(1995)}]{95-LAMMPS}
\bibinfo{author}{\bibfnamefont{S.}~\bibnamefont{Plimpton}},
  \bibinfo{journal}{J. Comput. Phys.} \textbf{\bibinfo{volume}{117}},
  \bibinfo{pages}{1 } (\bibinfo{year}{1995}).

\bibitem[{\citenamefont{Nosé}(1984)}]{84JCP-Nose}
\bibinfo{author}{\bibfnamefont{S.}~\bibnamefont{Nosé}}, \bibinfo{journal}{J.
  Chem. Phys.} \textbf{\bibinfo{volume}{81}}, \bibinfo{pages}{511}
  (\bibinfo{year}{1984}).

\bibitem[{\citenamefont{Hoover}(1985)}]{85PRA-Hoover}
\bibinfo{author}{\bibfnamefont{W.~G.} \bibnamefont{Hoover}},
  \bibinfo{journal}{Phys. Rev. A} \textbf{\bibinfo{volume}{31}},
  \bibinfo{pages}{1695} (\bibinfo{year}{1985}).

\bibitem[{\citenamefont{Martyna et~al.}(1994)\citenamefont{Martyna, Tobias, and
  Klein}}]{94JCP-MTK}
\bibinfo{author}{\bibfnamefont{G.~J.} \bibnamefont{Martyna}},
  \bibinfo{author}{\bibfnamefont{D.~J.} \bibnamefont{Tobias}},
  \bibnamefont{and} \bibinfo{author}{\bibfnamefont{M.~L.} \bibnamefont{Klein}},
  \bibinfo{journal}{J. Chem. Phys.} \textbf{\bibinfo{volume}{101}},
  \bibinfo{pages}{4177} (\bibinfo{year}{1994}).

\bibitem[{\citenamefont{Perdew et~al.}(1996)\citenamefont{Perdew, Burke, and
  Ernzerhof}}]{96L-PBE}
\bibinfo{author}{\bibfnamefont{J.~P.} \bibnamefont{Perdew}},
  \bibinfo{author}{\bibfnamefont{K.}~\bibnamefont{Burke}}, \bibnamefont{and}
  \bibinfo{author}{\bibfnamefont{M.}~\bibnamefont{Ernzerhof}},
  \bibinfo{journal}{Phys. Rev. Lett.} \textbf{\bibinfo{volume}{77}},
  \bibinfo{pages}{3865} (\bibinfo{year}{1996}).

\bibitem[{\citenamefont{Kittel}(1976)}]{Sn-exp-2}
\bibinfo{author}{\bibfnamefont{C.}~\bibnamefont{Kittel}},
  \emph{\bibinfo{title}{Introduction to Solid State Physics. Fifth edition}}
  (\bibinfo{publisher}{Wiley}, \bibinfo{year}{1976}).

\bibitem[{\citenamefont{Barrett}(1966)}]{Sn-exp-1}
\bibinfo{author}{\bibfnamefont{C.}~\bibnamefont{Barrett}},
  \emph{\bibinfo{title}{Structure of Metals; Crystallographic Methods,
  Principles, and Data}} (\bibinfo{publisher}{McGraw-Hill},
  \bibinfo{year}{1966}).

\bibitem[{\citenamefont{Gale and Totemeier}(2003)}]{Sn-exp-8}
\bibinfo{author}{\bibfnamefont{W.~F.} \bibnamefont{Gale}} \bibnamefont{and}
  \bibinfo{author}{\bibfnamefont{T.~C.} \bibnamefont{Totemeier}},
  \emph{\bibinfo{title}{Smithells metals reference book (Eighth Edition)}}
  (\bibinfo{publisher}{Butterworth-Heinemann}, \bibinfo{year}{2003}).

\bibitem[{\citenamefont{Ihm and Cohen}(1981)}]{Sn-exp-4}
\bibinfo{author}{\bibfnamefont{J.}~\bibnamefont{Ihm}} \bibnamefont{and}
  \bibinfo{author}{\bibfnamefont{M.~L.} \bibnamefont{Cohen}},
  \bibinfo{journal}{Phys. Rev. B} \textbf{\bibinfo{volume}{23}},
  \bibinfo{pages}{1576} (\bibinfo{year}{1981}).

\bibitem[{\citenamefont{Rayne and Chandrasekhar}(1960)}]{Sn-exp-3}
\bibinfo{author}{\bibfnamefont{J.~A.} \bibnamefont{Rayne}} \bibnamefont{and}
  \bibinfo{author}{\bibfnamefont{B.~S.} \bibnamefont{Chandrasekhar}},
  \bibinfo{journal}{Phys. Rev.} \textbf{\bibinfo{volume}{120}},
  \bibinfo{pages}{1658} (\bibinfo{year}{1960}).

\bibitem[{\citenamefont{Vaboya and Kennedy}(1970)}]{Sn-exp-7}
\bibinfo{author}{\bibfnamefont{S.}~\bibnamefont{Vaboya}} \bibnamefont{and}
  \bibinfo{author}{\bibfnamefont{G.}~\bibnamefont{Kennedy}},
  \bibinfo{journal}{J. Phys. Chem. Solids} \textbf{\bibinfo{volume}{31}},
  \bibinfo{pages}{2329 } (\bibinfo{year}{1970}).

\bibitem[{\citenamefont{Wang et~al.}(2020)\citenamefont{Wang, Geng, Wu, and
  Chen}}]{20JCP-Geng}
\bibinfo{author}{\bibfnamefont{Y.~X.} \bibnamefont{Wang}},
  \bibinfo{author}{\bibfnamefont{H.~Y.} \bibnamefont{Geng}},
  \bibinfo{author}{\bibfnamefont{Q.}~\bibnamefont{Wu}}, \bibnamefont{and}
  \bibinfo{author}{\bibfnamefont{X.~R.} \bibnamefont{Chen}},
  \bibinfo{journal}{J. Chem. Phys.} \textbf{\bibinfo{volume}{152}},
  \bibinfo{pages}{024118} (\bibinfo{year}{2020}).

\bibitem[{\citenamefont{Price et~al.}(1971)\citenamefont{Price, Rowe, and
  Nicklow}}]{71B-alpha_Sn}
\bibinfo{author}{\bibfnamefont{D.~L.} \bibnamefont{Price}},
  \bibinfo{author}{\bibfnamefont{J.~M.} \bibnamefont{Rowe}}, \bibnamefont{and}
  \bibinfo{author}{\bibfnamefont{R.~M.} \bibnamefont{Nicklow}},
  \bibinfo{journal}{Phys. Rev. B} \textbf{\bibinfo{volume}{3}},
  \bibinfo{pages}{1268} (\bibinfo{year}{1971}).

\bibitem[{\citenamefont{Rowe}(1967)}]{67PR-beta_Sn}
\bibinfo{author}{\bibfnamefont{J.~M.} \bibnamefont{Rowe}},
  \bibinfo{journal}{Phys. Rev.} \textbf{\bibinfo{volume}{163}},
  \bibinfo{pages}{547} (\bibinfo{year}{1967}).

\bibitem[{\citenamefont{Togo and Tanaka}(2015)}]{15SM-phonopy}
\bibinfo{author}{\bibfnamefont{A.}~\bibnamefont{Togo}} \bibnamefont{and}
  \bibinfo{author}{\bibfnamefont{I.}~\bibnamefont{Tanaka}},
  \bibinfo{journal}{Scr. Mater.} \textbf{\bibinfo{volume}{108}},
  \bibinfo{pages}{1} (\bibinfo{year}{2015}).

\bibitem[{\citenamefont{Itami et~al.}(2003)\citenamefont{Itami, Munejiri,
  Masaki, Aoki, Ishii, Kamiyama, Senda, Shimojo, and Hoshino}}]{03PRB-Itami}
\bibinfo{author}{\bibfnamefont{T.}~\bibnamefont{Itami}},
  \bibinfo{author}{\bibfnamefont{S.}~\bibnamefont{Munejiri}},
  \bibinfo{author}{\bibfnamefont{T.}~\bibnamefont{Masaki}},
  \bibinfo{author}{\bibfnamefont{H.}~\bibnamefont{Aoki}},
  \bibinfo{author}{\bibfnamefont{Y.}~\bibnamefont{Ishii}},
  \bibinfo{author}{\bibfnamefont{T.}~\bibnamefont{Kamiyama}},
  \bibinfo{author}{\bibfnamefont{Y.}~\bibnamefont{Senda}},
  \bibinfo{author}{\bibfnamefont{F.}~\bibnamefont{Shimojo}}, \bibnamefont{and}
  \bibinfo{author}{\bibfnamefont{K.}~\bibnamefont{Hoshino}},
  \bibinfo{journal}{Phys. Rev. B} \textbf{\bibinfo{volume}{67}},
  \bibinfo{pages}{064201} (\bibinfo{year}{2003}).

\bibitem[{\citenamefont{Assael et~al.}(2010)\citenamefont{Assael, Kalyva,
  Antoniadis, Michael~Banish, Egry, Wu, Kaschnitz, and Wakeham}}]{10JPC-Assael}
\bibinfo{author}{\bibfnamefont{M.~J.} \bibnamefont{Assael}},
  \bibinfo{author}{\bibfnamefont{A.~E.} \bibnamefont{Kalyva}},
  \bibinfo{author}{\bibfnamefont{K.~D.} \bibnamefont{Antoniadis}},
  \bibinfo{author}{\bibfnamefont{R.}~\bibnamefont{Michael~Banish}},
  \bibinfo{author}{\bibfnamefont{I.}~\bibnamefont{Egry}},
  \bibinfo{author}{\bibfnamefont{J.}~\bibnamefont{Wu}},
  \bibinfo{author}{\bibfnamefont{E.}~\bibnamefont{Kaschnitz}},
  \bibnamefont{and} \bibinfo{author}{\bibfnamefont{W.~A.}
  \bibnamefont{Wakeham}}, \bibinfo{journal}{J. Phys. Chem. Ref. Data}
  \textbf{\bibinfo{volume}{39}}, \bibinfo{pages}{033105}
  (\bibinfo{year}{2010}).

\bibitem[{\citenamefont{Einstein}(1905)}]{1905AP-Einstein}
\bibinfo{author}{\bibfnamefont{A.}~\bibnamefont{Einstein}},
  \bibinfo{journal}{Ann. Phys.} \textbf{\bibinfo{volume}{322}},
  \bibinfo{pages}{549} (\bibinfo{year}{1905}).

\bibitem[{\citenamefont{Van~der Ven et~al.}(2001)\citenamefont{Van~der Ven,
  Ceder, Asta, and Tepesch}}]{01B-Van}
\bibinfo{author}{\bibfnamefont{A.}~\bibnamefont{Van~der Ven}},
  \bibinfo{author}{\bibfnamefont{G.}~\bibnamefont{Ceder}},
  \bibinfo{author}{\bibfnamefont{M.}~\bibnamefont{Asta}}, \bibnamefont{and}
  \bibinfo{author}{\bibfnamefont{P.~D.} \bibnamefont{Tepesch}},
  \bibinfo{journal}{Phys. Rev. B} \textbf{\bibinfo{volume}{64}},
  \bibinfo{pages}{184307} (\bibinfo{year}{2001}).

\bibitem[{\citenamefont{Marcolongo and Marzari}(2017)}]{17PRM-Marcolongo}
\bibinfo{author}{\bibfnamefont{A.}~\bibnamefont{Marcolongo}} \bibnamefont{and}
  \bibinfo{author}{\bibfnamefont{N.}~\bibnamefont{Marzari}},
  \bibinfo{journal}{Phys. Rev. Mater.} \textbf{\bibinfo{volume}{1}},
  \bibinfo{pages}{025402} (\bibinfo{year}{2017}).

\bibitem[{\citenamefont{Yeh and Hummer}(2004)}]{04JPCB-Yeh}
\bibinfo{author}{\bibfnamefont{I.-C.} \bibnamefont{Yeh}} \bibnamefont{and}
  \bibinfo{author}{\bibfnamefont{G.}~\bibnamefont{Hummer}},
  \bibinfo{journal}{J. Phys. Chem. B} \textbf{\bibinfo{volume}{108}},
  \bibinfo{pages}{15873} (\bibinfo{year}{2004}).

\bibitem[{\citenamefont{Bruson and Gerl}(1980)}]{80PRB-Bruson}
\bibinfo{author}{\bibfnamefont{A.}~\bibnamefont{Bruson}} \bibnamefont{and}
  \bibinfo{author}{\bibfnamefont{M.}~\bibnamefont{Gerl}},
  \bibinfo{journal}{Phys. Rev. B} \textbf{\bibinfo{volume}{21}},
  \bibinfo{pages}{5447} (\bibinfo{year}{1980}).

\bibitem[{\citenamefont{Onishi et~al.}(1998)\citenamefont{Onishi, Inatomi,
  Tanaka, Shinozaki, Watanabe, Fujimoto, and Itoh}}]{98JJSMA-onishi}
\bibinfo{author}{\bibfnamefont{F.}~\bibnamefont{Onishi}},
  \bibinfo{author}{\bibfnamefont{Y.}~\bibnamefont{Inatomi}},
  \bibinfo{author}{\bibfnamefont{T.}~\bibnamefont{Tanaka}},
  \bibinfo{author}{\bibfnamefont{N.}~\bibnamefont{Shinozaki}},
  \bibinfo{author}{\bibfnamefont{M.}~\bibnamefont{Watanabe}},
  \bibinfo{author}{\bibfnamefont{A.}~\bibnamefont{Fujimoto}}, \bibnamefont{and}
  \bibinfo{author}{\bibfnamefont{K.}~\bibnamefont{Itoh}}, \bibinfo{journal}{J.
  Jpn. Soc. Microgravity Appl.} \textbf{\bibinfo{volume}{15}},
  \bibinfo{pages}{225} (\bibinfo{year}{1998}).

\bibitem[{\citenamefont{Vega et~al.}(2008)\citenamefont{Vega, Sanz, Abascal,
  and Noya}}]{08JPCM-Vega}
\bibinfo{author}{\bibfnamefont{C.}~\bibnamefont{Vega}},
  \bibinfo{author}{\bibfnamefont{E.}~\bibnamefont{Sanz}},
  \bibinfo{author}{\bibfnamefont{J.~L.~F.} \bibnamefont{Abascal}},
  \bibnamefont{and} \bibinfo{author}{\bibfnamefont{E.~G.} \bibnamefont{Noya}},
  \bibinfo{journal}{J. Phys. Condens. Matter} \textbf{\bibinfo{volume}{20}},
  \bibinfo{pages}{153101} (\bibinfo{year}{2008}).

\bibitem[{\citenamefont{Rehn et~al.}(2021)\citenamefont{Rehn, Greeff,
  Burakovsky, Sheppard, and Crockett}}]{21B-Rehn}
\bibinfo{author}{\bibfnamefont{D.~A.} \bibnamefont{Rehn}},
  \bibinfo{author}{\bibfnamefont{C.~W.} \bibnamefont{Greeff}},
  \bibinfo{author}{\bibfnamefont{L.}~\bibnamefont{Burakovsky}},
  \bibinfo{author}{\bibfnamefont{D.~G.} \bibnamefont{Sheppard}},
  \bibnamefont{and} \bibinfo{author}{\bibfnamefont{S.~D.}
  \bibnamefont{Crockett}}, \bibinfo{journal}{Phys. Rev. B}
  \textbf{\bibinfo{volume}{103}}, \bibinfo{pages}{184102}
  (\bibinfo{year}{2021}).

\bibitem[{\citenamefont{Armiento and Mattsson}(2005)}]{05B-AM05}
\bibinfo{author}{\bibfnamefont{R.}~\bibnamefont{Armiento}} \bibnamefont{and}
  \bibinfo{author}{\bibfnamefont{A.~E.} \bibnamefont{Mattsson}},
  \bibinfo{journal}{Phys. Rev. B} \textbf{\bibinfo{volume}{72}},
  \bibinfo{pages}{085108} (\bibinfo{year}{2005}).

\end{thebibliography}

\end{document}